\documentclass[journal,comsoc]{IEEEtran}
\usepackage{amsmath,amsfonts}
\usepackage{algorithmic}
\usepackage{array}
\usepackage[caption=false,font=normalsize,labelfont=sf,textfont=sf]{subfig}
\usepackage{textcomp}
\usepackage{stfloats}
\usepackage{url}
\usepackage{verbatim}
\usepackage{graphicx}
\def\BibTeX{{\rm B\kern-.05em{\sc i\kern-.025em b}\kern-.08em
    T\kern-.1667em\lower.7ex\hbox{E}\kern-.125emX}}
\usepackage{balance}
 \usepackage{array}
\usepackage{balance}
\usepackage{boldline}
\usepackage{calc}
\usepackage{dcolumn}
\usepackage{diagbox}
\usepackage{enumitem}
\usepackage{graphicx}
\usepackage{hhline}
\usepackage{listings}
\usepackage{makecell}
\usepackage{multirow}
\usepackage{setspace}
\usepackage{tabularx}
\usepackage{url}
\usepackage{xcolor}
\usepackage{xspace}
\usepackage{adjustbox}

\date{}  

\let\oldtabular\tabular
\renewcommand{\tabular}{\footnotesize\oldtabular}

\let\oldtabularx\tabularx
\renewcommand{\tabularx}{\footnotesize\oldtabularx}

\newcommand{\name}{\textit{NetDx}\xspace}
\newcommand{\mbnd}{\textit{MBND}\xspace}
\newcommand{\emuname}{\textit{Podnet}\xspace}

\newcolumntype{d}[1]{D{.}{.}{#1}}

\newcommand{\ctxtlr}[1]{\multicolumn{1}{|c|}{#1}}
\newcommand{\ctxtl}[1]{\multicolumn{1}{|c}{#1}}

\newcommand{\vcentrows}[1]{\multirow{2}{*}{#1}}

  \usepackage[absolute,overlay]{textpos}
\begin{document}

\title{Model-Based Diagnosis: Automating End-to-End
Diagnosis of Network Failures}

\author{
Changrong Wu,
Yiyao Yu,
Myungjin Lee,
Jayanth Srinivasa,
Ennan Zhai,
George Varghese,
Yuval Tamir\thanks{Changrong Wu was with UCLA and is now with Microsoft.}
\thanks{Yiyao Yu, George Varghese, and Yuval Tamir
are with UCLA.}
\thanks{Myungjin Lee and Jayanth Srinivasa are with
Cisco Systems.}
\thanks{Ennan Zhai is with Alibaba Cloud.}}

\maketitle

\begin{textblock*}{\textwidth}[0.5,0](0.5\paperwidth,1.95in) \centering
July 2025
\end{textblock*}

\begin{abstract}
Fast diagnosis and repair of enterprise network failures
is critically important since disruptions cause major business impacts.
Prior works focused on diagnosis
primitives or procedures limited to a subset of
the problem, such as only data plane or only control plane faults.
This paper proposes a new paradigm,
\textit{model-based network diagnosis}, that
provides a \textit{systematic} way to derive 
\textit{automated} procedures for identifying the
\textit{root cause}
of network failures, based on reports of end-to-end
user-level symptoms.
The diagnosis procedures are
\textit{systematically derived} from a model
of packet forwarding and routing,
covering hardware, firmware, and software faults in
both the data plane and distributed control plane.
These automated procedures replace and dramatically accelerate
diagnosis by an experienced human operator.
Model-based diagnosis is inspired by, leverages, and is
complementary to recent work on network verification.
We have built \textit{NetDx}, a proof-of-concept implementation
of model-based network diagnosis.
We deployed \textit{NetDx} on a new emulator of networks
consisting of P4 switches with distributed routing software.
We validated the robustness and coverage
of \textit{NetDx} with an automated fault injection campaign,
in which 100\% of faults were diagnosed correctly.
Furthermore, on a data set of 33 faults
from a large cloud provider that are
within the domain targeted by \textit{NetDx},
30 are efficiently diagnosed in seconds instead of hours.
 \end{abstract}

\section{Introduction}
\label{sec:intro}

Failures in large operational enterprise networks and clouds are often
subtle and can take hours to diagnose.
This affects their ability to
provide four nines availability, which allows
only 52 minutes of down time in a year~\cite{awsreliability}.
In practice, the diagnosis process often requires
an expert human operator manually deploying
multiple tools, such as traceroute and Wireshark;
a process that is ad hoc, error prone, and can
take multiple hours.
In many cases, the complexity of the diagnosis task is due to
a semantic (data plane vs. control plane)
and physical distance between the observed symptom
and the root cause of the failure.

Past relevant research focused on diagnosis primitives
either for data plane faults
(often in the context of SDN)~\cite{handigol2014netsight,
zhu2015everflow, handigol2012ndb, jeyakumar2014minions,
tammana2016pathdump, tammana2018switchpointer}
or for control plane faults~\cite{lad2004linkrank,
colitti2005bgplay, blazakis2006bgp-inspect, cittadini2008bgpath,
haeberlen2009netreview},
without any way to automatically
identify the root cause of the failure, taking into
account interactions between the data and control planes.
In contrast, this paper introduces a new
top-down model-based paradigm that is completely different from
past work (\S\ref{sec:related}).
Crucially, this paradigm facilitates handling
of complex data/control plane interactions, \textit{automating}
the procedures of an experienced network operator, improving
failure response times
\textbf{from hours to seconds}
(\S\ref{subsec:campaign_results}).

Recent work in network verification~\cite{beckett2017minesweeper,
fogel2015batfish, HSA, ERA, NoD, ye2020hoyan, tian2019jinjing}
has exploited the fact that the correct network
behavior is defined by network protocols, the network topology,
and switch configurations.
This allows the construction of a simple end-to-end
model of packet forwarding and routing, and the use
of this model to identify configuration faults.

In this paper, we apply this insight
of the essential \textit{simplicity} of networks
for a different purpose.
We propose \textit{model-based network diagnosis}, in which
network failures caused by switch or link faults can be diagnosed
based on deviations from a formal end-to-end model of packet
forwarding and routing (\S\ref{sec:prim_proc}).
It should be noted that the term
``model-based diagnosis'' has been used
in other contexts~\cite{de2003MBDfundamentals,
darwiche2000model, de1999model} with a different meaning.

\textit{Model-based network diagnosis}
is different from network verification
since the goal is to identify faults in an operational
network, not errors in configurations.
Unlike model-based verification, that can be done
proactively, switch or link failures happen in real-time
and the resulting network failures must be responded to
quickly in the field, using cues from the operational network.
A global cloud provider with which we are working
reports that, due to configuration verification tools,
configuration errors have been greatly reduced and
66\% of their current
network failures are now due to switch hardware or software faults.

Switch or link failures may be caused by hardware faults.
Switch failures may also be caused by software or firmware faults.
We target diagnosis of permanent or high-frequency
intermittent faults that ultimately result
in excessive data packet drops or corruption.
It should be noted that
in the set of real world failures that
we collected (\S\ref{subsec:fault_set}), the overwhelming
majority of hardware or software faults were permanent and
resulted in data packet drops or corruption.

\begin{table*}
    \centering
    \caption{Unlike prior work, model-based diagnosis allows
    automated diagnosis of complex data plane/control plane faults.}
    \label{tab:related_work}
    \begin{tabular}{| @{ }p{1.25in}|p{1.4in}|p{0.6in}|@{ }c@{ }|c|c|c|}
    \hline
    \ctxtl{\vcentrows{Category}} &
       \ctxtlr{\vcentrows{Canonical Example(s)}} &
    \parbox[t][][t]{\widthof{Automateda}}{
       \setstretch{0.8} Automated \\ Diagnosis?} &
    \multirow{2}{*}{Overhead} &
    \parbox[t][][t]{\widthof{Control Plane/Data Plane}}
       {\setstretch{0.8} Data Plane/Control Plane \\ Interactions} &
    \parbox[t][][t]{0.5in}{\setstretch{0.8} Diagnosis \\ Granularity} &
    \parbox[t][][t]{0.5in}{\setstretch{0.8} Performance \\ Faults?} \\
    \hlineB{2.5}
    Monitoring & NetBouncer\cite{NetBouncer}, Trumpet\cite{moshref2016trumpet} & Yes & low & No & Switch/Link & Yes \\ 
    \hline
    New Switch Primitives & FlowRadar\cite{flowradar}, NDB\cite{handigol2012ndb} & No & High & No & Switch/Link & Yes \\
    \hline
    HeaderMods + Primitives & INT\cite{INT}, PathDump\cite{tammana2016pathdump} & No & High & No & Switch/Link & Yes \\
    \hline
    Data Plane Log Analysis & Everflow\cite{zhu2015everflow} & No & High & No or limited & Switch/Link & Yes \\
    \hline
    Control Plane Log Analysis & BGP-Inspect\cite{blazakis2006bgp-inspect}, BGPlay\cite{colitti2005bgplay} & No & High & No & Switch/ISP & No \\
    \hline
    \vcentrows{Higher Level Abstractions} &
       \parbox[t][][t]{1.25in}{\setstretch{0.9}
	  dShark\cite{dShark}, Marple\cite{narayana2017marple},
	  \newline Everflow\cite{zhu2015everflow}} &
       \parbox[t][][t]{0.6in}{\setstretch{0.9}
       No but raises \\ abstraction} &
       \vcentrows{High} & \vcentrows{No} & \vcentrows{Switch} &
       \vcentrows{Yes} \\
    \hlineB{2.0}
    \textbf{Model-based Diagnosis} & \textbf{NetDx}~[this paper] &
    \textbf{Yes} & \textbf{Low} & \textbf{Yes} &
    \textbf{Switch/Link} & \textbf{No}\\
    \hlineB{2.0}
\end{tabular}
 \end{table*}

We use model-based diagnosis to design and implement
a system, called \name,
that takes as input an end-to-end \textit{failure report}
and \textit{automatically} identifies the root cause
failed switch or link.
A human user or an application generates the failure report
that typically describes a loss of
connection between two hosts or groups of
hosts for all packets or some subset of packets.
The goal of the diagnosis is to help network operators quickly
resolve network disruptions.
Thus, the diagnosis process only needs to identify the
faulty \textit{switch} or \textit{link} rather than determine exactly
which internal switch component has failed.

Based on a formal model of the \textit{functionality}
of network switches, we derive a fault model that
describes how switch functionality can be disrupted.
These models define the correct and faulty behaviors 
of the switch \textit{at its interfaces} to the rest
of the network and
are \textit{independent} of implementation details.
Using these models and the specification of network protocols,
we \textit{systematically} derive
\textit{automated} diagnosis procedures by
tracing the steps of packet forwarding and route propagation.
At each step, the actions of switches in the operational
network are compared to the correct actions derived
from the model and the comparison results
guide the diagnosis process.
We leverage a configuration analysis tool~\cite{fogel2015batfish}
to provide a representation of the network model
that is directly comparable to what is observable
in the operational network.

A key challenge in the implementation of \name is
efficient collection of
information from the operational
network in order to compare it to the model.
Many proposed network diagnosis mechanisms rely on
packet mirroring and logging to servers
(e.g., \cite{zhu2015everflow,dShark}),
potentially requiring significant overhead in terms
of network bandwidth and host resources.
Some mechanisms piggyback diagnostic
information on packets, potentially leading to
packet fragmentation and resource usage at the hosts
for processing the piggybacked information.
In contrast,
to minimize overhead, \name{}'s mechanisms are
based on data collection primitives implemented in the switches
that require minimal processing and
storage resources on the switches and essentially
no overhead in terms of
network bandwidth and host resources (\S\ref{subsec:dxprimitives}).

Table~\ref{tab:related_work} compares \name to prior work.
While the detailed explanation of the Table is left
to \S\ref{sec:related}, the main takeaways are as follows.
The current model is limited to IP/BGP networks
and only deals with faults that manifest as packet drops
or packet corruption.
\name does not yet handle performance faults or some key features
(e.g., overlays networks).
However, the current \name implementation still accomplishes
a task that has not been demonstrated previously:
automatically identifying the root cause of complex failure
scenarios that involve the
interaction of the data and control planes,
improving failure
response times \textit{from hours to seconds}
(\S\ref{subsec:campaign_results}).
We believe that our new model-based paradigm
is extensible in a straightforward manner to, for example, other
routing protocols and overlay networks.
In contrast, for many real failure scenarios,
the primitives introduced by prior work still
need to be combined manually without a model to guide automation.

To evaluate \name, we built a network emulator
that executes real routing software (FRR),
injects realistic faults, and allows us
to implement and test
a \name implementation that is directly-transferable
to a physical network.
To enable modification of the data plane,
the data plane of
each emulated switch is an instance of the P4~\cite{p4paper}
behavioral simulator~\cite{bmv2}.
Thus, the data plane is programmable.

We collected from a large cloud provider
33 real-world network failure scenarios
(\S\ref{subsec:fault_set})
that are within the domain targeted by \name (\S\ref{sec:model})
that occurred over several years.
Their diagnosis was originally done manually and, in most cases,
took multiple hours (median of 4.5 hours).
We used these scenarios as a sanity check on our
fault model and diagnosis procedures (\S\ref{sec:netdx_diag})
as well as to drive some of our experiments with \name.
The evaluation of \name included an automated
random fault injection campaign in which
\name was able to correctly diagnose all injected faults.

\noindent
{\bf Contributions:}
1)~Introduction of the novel, implementation-agnostic paradigm of
\textit{model-based network diagnosis} that
enables \textit{automated,} \textit{end-to-end}
diagnosis of operational enterprise networks,
taking into account both data plane and control plane faults.
2)~Derivation of a functional fault model
of network switches
that is completely independent of implementation details
and can be used by others to evaluate network diagnosis
mechanisms.
3)~Design and implementation of a diagnosis system, \name,
based on data collection at the switches,
that is \textit{systematically derived} from the network model,
leverages the representation of the model by a
configuration analysis tool (Batfish~\cite{fogel2015batfish}),
does not intrude on normal network operations,
and only requires a small set of primitives and
few resources.
4)~Evaluation of \name using an
automated fault injection campaign
and a small user study comparing to manual diagnosis.
 \section{Preliminary Discussion}
\label{sec:prelim}

To dispel potential confusion, this section clarifies
terminology and key concepts
central to \textit{model-based network diagnosis}
(henceforth \mbnd) and \name.

\noindent
\textbf{\mbnd versus \name:}
\name (\S\ref{sec:netdx_diag}, \S\ref{sec:implementation})
is a proof-of-concept implementation
of \mbnd (\S\ref{sec:model}, \S\ref{sec:prim_proc}).
However, other implementations of \mbnd are possible.
Starting from a report of end-to-end (host-to-host) user-level
symptoms,
\mbnd guides the derivation of \textit{what} information
needs to be collected and the order in which the information
should be collected to identify the \textit{root cause}
of the particular network failure.
\mbnd does not determine \textit{how} to collect this information.
\name shows one possible efficient implementation of
how to collect the information and use it, following
procedures derived using \mbnd, to reach diagnosis.

\noindent
\textbf{The meaning of \textit{root-cause diagnosis}:}
In this paper, \textit{root-cause diagnosis} means identifying
the switch or link that is actually faulty.
This is different from the first location where a symptom
of the fault is detected.
For example, packets may be unexpectedly dropped by a fault-free
switch, \textit{SWA}, since another fault-free
switch, \textit{SWB}, forwards the packets in the wrong direction
due to an incorrect route advertisement sent
by a faulty switch, \textit{SWC}.
\textit{Root-cause diagnosis} identifies the problem to
be \textit{SWC}, not \textit{SWA} or \textit{SWB}.
Once the actual faulty switch is identified, it can be 
quickly replaced or configured out for separate offline
detailed diagnosis by the switch manufacturer.

\noindent
\textbf{Diagnosis versus detection:}
Network failure \textit{detection} mechanisms simply
determine whether there is a fault somewhere in the network.
On the other hand, the goal of a \textit{diagnosis} mechanism
(\mbnd and its implementation, such as \name) is to identify the 
faulty switch or link.

\noindent
\textbf{Interactions between the data plane and control plane
must be accounted for:}
As stated earlier, the complexity of the diagnosis task is due to
a semantic and physical distance between the observed symptom
and the root cause of the failure.
Consider the example described above.
In such a scenario,
existing techniques, such as ATPG~\cite{ATPG}, can identify
the switch where packets are dropped (\textit{SWA}), but not the
actual faulty switch (\textit{SWC}) that generates the incorrect
route advertisement.
A high-overhead mechanism that examines every FIB
in the network may be able to identify \textit{SWB},
but not \textit{SWC}.

\noindent
\textbf{The meaning of \textit{end-to-end diagnosis}:}
A network failure is manifested as
a problem in the communication among hosts.
\textit{End-to-end diagnosis} is triggered
by such a failure report and identifies the
root-cause of the failure.
Based on \mbnd, such diagnosis can be done \textit{automatically}.

\noindent
\textbf{The meaning of \textit{automated} diagnosis:}
\mbnd enables and guides systematic development of
diagnosis procedures by human experts.
These procedures are developed once and can then be used
numerous times
in multiple locations, without experts,
to quickly and autonomously diagnose network failures.

 \section{A Model of Network Switches}
\label{sec:model}

\begin{table*}[t]
    \centering
    \captionsetup{justification=centering}
    \caption{\setstretch{0.9} A functional model of fault-free and
       faulty switches. \\
       Faulty functionality (the fault model)
	     is systematically derived by negating
	     the fault-free functionality.}
    \label{tab:formal_model}
    {\setlength{\tabcolsep}{3pt}
\begin{tabular}{V{3.0} @{ } c @{ } V{3.0}  p{6.0in} V{3.0} @{} c}
  \clineB{1-2}{3.0}
  \multirow{10}{*}{\rotatebox{90}{Packet Forwarding}} &
   The switch forwards packets with characteristics $C_{forward}$
      to a port in $Q$ when the state of S meets condition
      $S_{uncongested}$ \newline
   The switch drops packets with characteristics $C_{forward}$
      when the state of S meets condition $S_{congested}$ \newline
   The switch drops packets with characteristics
      $\overline{C_{forward}}$ regardless of the state \\
  \cline{2-2}
  & $EgressPort(C_{forward}, S_{uncongested}) = p, p \in Q,
     Q \subset P$ \newline
  $EgressPort(C_{forward}, S_{congested}) = \text{Null}$ \hspace{1.0in}
  $EgressPort(\overline{C_{forward}}, *) = \text{Null}$ \\
  \clineB{2-3}{2.0}
  &  \parbox{5.5in}{\vspace{2pt}
     Switch S drops packets with characteristics $C_{forward}$
     when the state of the switch meets condition $S_{uncongested}$
     \textbf{\textit{or}}} &
     \multirow{5}{*}{
           \rotatebox{90}{\textbf{FAULTY}} } \\
  & Switch S forwards packets with characteristics $C_{forward}$
    to a port $p, p \notin Q$ \textbf{\textit{or}}\\
  & Switch S forwards packets with characteristics
     $\overline{C_{forward}}$ to a port in the switch's port set $P$ \\
  \cline{2-2}
  & $EgressPort(C_{forward}, S_{uncongested}) = Null$\\
  & $EgressPort(C_{forward}, *) = p, p \notin Q$ \hspace{1.17in}
    $EgressPort(\overline{C_{forward}}, *) = p', p' \in P$ \\
  \hlineB{3.0}
  \multirow{8}{*}{\rotatebox{90}
     {\parbox{\widthof{Transformation}}{
        \setstretch{0.8} Packet \\ Transformation}}} &
  \parbox{5.5in}{\vspace{1pt}The switch transforms the header of packet Pkt(C) with
     characteristics $C$ by function Fhdrt} \\
  & The switch should not change the payload of the packet \\
  \cline{2-2}
  & $NewPkt = Fhdrt(Pkt(C))$ \\
  & $NewPkt.payload = Pkt(C).payload$ \\
  \clineB{2-3}{2.0}
  & \parbox{5.5in}{\vspace{2pt}
    The switch transforms the header of packet Pkt(C) with
     characteristics $C$ by function $Fbadt$, $Fbadt \neq Fhdrt$} &
     \multirow{4}{*}{\rotatebox{90}{\textbf{FAULTY }} } \\
  & The switch changes the payload of packet Pkt(C)\\
  \cline{2-2}
  & $NewPkt = Fbadt(Pkt(C)) \neq Fhdrt(Pkt(C))$ \\
  & $NewPkt.payload \neq Pkt(C).payload$ \\
  \hlineB{3.0}
  \multirow{6}{*}{\rotatebox{90}
     {\parbox{\widthof{\fontsize{7}{7}\selectfont Data Plane }}{
        \raggedright\fontsize{7}{6}\selectfont
         Data Plane \\ Table \\ Generation}}} &
   The switch generates data plane tables according to their
      corresponding control plane tables \\
  \cline{2-2}
  & \parbox{5.5in}{\vspace{2pt}
     $\forall Entry \in ControlPlaneTable,
     \exists Entry' \in DataPlaneTable
     \ s.t. \ Entry \equiv Entry' \ \land$ \newline
    $\forall Entry \in DataPlaneTable,
       \exists Entry' \in ControlPlaneTable
       \ s.t. \ Entry \equiv Entry'$} \\
  \clineB{2-3}{2.0}
  & \parbox{5.55in}{\vspace{2pt}
    One of switch S's data plane tables is inconsistent with
     the corresponding control plane table} &
     \multirow{3}{*}{\rotatebox{90}{
          \textbf{\fontsize{6.5}{6.5}\selectfont FAULTY }}}  \\
  \cline{2-2}
  & \parbox{5.5in}{\vspace{2pt}
     $\exists Entry \in ControlPlaneTable,
     \forall Entry' \in
     DataPlaneTable, Entry' \not\equiv Entry \ \lor$ \newline
  $\exists Entry \in DataPlaneTable,
     \forall Entry' \in ControlPlaneTable, Entry' \not\equiv Entry$} \\
  \hlineB{3.0}
  \multirow{4}{*}{\rotatebox{90}
     {\parbox{\widthof{\fontsize{7}{7}\selectfont Generation}}{
	\raggedright\fontsize{7}{6}\selectfont
        Route \\ Table \\ Generation}}} &
     The switch computes Route Table using specified routing
        algorithms based on configuration and received
	routing information \\
  \cline{2-2}
  & $RouteTable = RouteCompute(Configuration,
     RecvRoutingInformation)$ \\
  \clineB{2-3}{2.0}
  &  \parbox{5.55in}{\vspace{2pt}
     The switch's Route Table is inconsistent with the result computed
     from the configuration and received routing information} &
     \multirow{2}{*}{\rotatebox{90}{
        \textbf{\fontsize{6.5}{6.5}\selectfont FLTY }}} \\
  \cline{2-2}
  & $RouteTable \neq RouteCompute(Configuration,
     RecvRoutingInformation)$ \\
  \hlineB{3.0}
  \multirow{4.5}{*}{\rotatebox{90}
     {\parbox{\widthof{\fontsize{7}{9}\selectfont vertizement}}{
	\raggedright\fontsize{7}{6}\selectfont
        Route Advertisement Reception}}} &
    The switch computes received routing information based on
      configuration and inbound route advertisements \\
  \cline{2-2}
  & $RecvRoutingInformation = RouteAdvReception(Configuration,
     InboundRouteAdv)$ \\
  \clineB{2-3}{2.0}
  &  \parbox{5.95in}{\setstretch{0.80}\raggedright\vspace{3pt}
     Switch's received routing information is inconsistent with
     the result computed from configuration and inbound route
     advertisements\vspace{2pt}} &
     \multirow{2}{*}{\rotatebox{90}{
        \textbf{\fontsize{6.5}{6.5}\selectfont FLTY }}} \\
  \cline{2-2}
  & $RecvRoutingInformation \neq
     RouteAdvReception(Configuration, InboundRouteAdv)$ \\
  \hlineB{3.0}
  \multirow{4.5}{*}{\rotatebox{90}
     {\parbox{\widthof{\fontsize{7}{9}\selectfont vertizement}}{
    \raggedright\fontsize{7}{6}\selectfont
    Route Advertisement Generation}}} &
    The switch generates outbound route advertisements based on
       configuration and Route Table \\
  \cline{2-2}
  & $OutboundRouteAdv =
     RouteAdvGeneration(Configuration, RouteTable)$ \\
  \clineB{2-3}{2.0}
  & \parbox{5.5in}{\vspace{2pt}
    The switch's outbound route advertisements are inconsistent
     with the result computed from configuration and Route Table} &
     \multirow{2}{*}{\rotatebox{90}{
        \textbf{\fontsize{6.5}{6.5}\selectfont FLTY }}} \\
  \cline{2-2}
  & $OutboundRouteAdv \neq
     RouteAdvGeneration(Configuration, RouteTable)$ \\
  \hlineB{3.0}
  \multirow{5}{*}{\rotatebox{90}
     {\parbox{\widthof{\fontsize{7}{7}\selectfont with External}}{
        \raggedright\fontsize{7}{6}\selectfont
        Interaction with External Entities}}} &
    The switch responds to messages from external entities
       following protocols and configurations \\
  \cline{2-2}
  & $Response(InboundMessage, Configuration) = ProtocolMessage$ \\
  \clineB{2-3}{2.0}
  & \parbox{5.5in}{\vspace{2pt}
     The switch responds to messages from external entities with
     Incorrect messages or it has no response to messages} &
     \multirow{3}{*}{\rotatebox{90}{
        \textbf{\fontsize{6.5}{6.5}\selectfont FAULTY }}} \\
  \cline{2-2}
  & $Response(InboundMessage, Configuration) \neq
     ProtocolMessage \ \lor$ \newline
    $Response(InboundMessage, Configuration) = Null$\\
  \hlineB{3.0}
\end{tabular}
}
 \end{table*}

\begin{table}[h]
    \centering
    \caption{The notation used in Table~\ref{tab:formal_model}.}
    \label{tab:notation}
    \newcommand{\txtlr}[1]{\multicolumn{1}{|c|}{#1}}
\newcommand{\txtr}[1]{\multicolumn{1}{c|}{#1}}

\begin{tabular}{| @{ } p{0.9in}|p{1.9in}|}
   \hline
   \txtlr{\textbf{Symbol}} & \txtr{\textbf{Semantics}} \\
   \hlineB{2.5}
   $S$ & A state of the switch's packet buffer\\
   \hline
   $P$ & The set of all the switch's interfaces\\
   \hline
   $Q$ & A set of switch interfaces, $Q \subset P$\\
   \hline
   $C$ & A set of packet characteristics\\
   \hline
   \textit{Fhdrt} & Packet header transformation function\\
   \hline
   \textit{DataPlaneTable} & Forwarding \& filtering data plane tables\\
   \hline
   \textit{ControlPlaneTable} & Routing \& filtering control plane tables\\
   \hline
\end{tabular}
 \end{table}

This section presents a functional model of network switches
which is then used to derive a switch fault model.
These are the first and second steps in model-based diagnosis.
The next section explains the third and fourth steps
in which these two models are used to derive 
diagnosis primitives and automated diagnosis procedures.

Since the required resolution of diagnosis is a switch or link,
diagnosis is only concerned with the \textit{functionality}
of these components at their interfaces to the rest of the network.
While network switches are complex, their behavior
\textit{at their interfaces} is defined by
their configurations and standard network protocols.
This behavior forms the functional model of
the switches that is \textit{independent} of low-level
implementation details.
A fault in a switch is manifested as a deviation
from the functionality defined by the switch model.

Model-based diagnosis can potentially deal
with a wide variety of faults, including those that
cause performance anomalies.
However, since this work only deals with network failures
that manifest as end-to-end (host-to-host) packet drops,
the relevant switch model only needs to capture forwarding
and routing.
Thus, our model assumes that every switch includes
structures, referred to as \textit{data plane tables},
that are involved with the handling of every packet.
The contents of the data plane tables are derived
by the switch from other structures on the switch, referred to
as \textit{control plane table}.
Examples of data plane tables are structures
that permit or deny reception
of packets (e.g., ACL tables) and
guide the forwarding of packets (e.g., FIBs).
Examples of control plane tables are structures that
contains routing information (e.g., RIBs)
as well as the configuration of packet permit/deny criteria.

Table~\ref{tab:formal_model} presents the high-level
functional model of switches as well as the switch
fault model \textit{derived} from the functional model.
The data plane portion is similar to functional
models used in data plane verification,
like Anteater~\cite{anteater} and HSA~\cite{HSA},
while the combination of data and control plane models
is most similar to Minesweeper~\cite{minesweeper}
and  ERA~\cite{ERA}.
However, the work in network verification does not consider negations
of the functional model to create a fault model.

The model consists of seven functionality categories.
For each category there are four rows, where the top
two rows contain informal and formal definitions
of the functionality of a fault-free switch.
The bottom two rows for each category contain
the informal and formal definitions of the
functionality of a faulty switch.
The definition of the functionality
of a faulty switch is derived by negating
the functionality of a fault-free switch.
The list of faulty functionalities forms the switch
\textit{fault model}.
Table~\ref{tab:notation} is a key to the notation.

As an example,
the top category in Table~\ref{tab:formal_model} defines
the packet forwarding function of a switch.
Specifically, a fault-free switch forwards a packet
to one of a subset of the switch's interfaces.
This forwarding is performed only if the packet header
characteristics meet configured criteria ($C_{forward}$)
as long as it does not encounter a full packet buffer
due to congestion (the buffer state is $S_{uncongested}$).
The packet is correctly dropped if
the header characteristics are in $\overline{C_{forward}}$, i.e.,
they do not meet the
configured forwarding criteria $C_{forward}$.
Furthermore, the model covers the possibility
that, in a fault-free switch, a packet may be correctly
dropped if it encounters
a full packet buffer ($S_{congested}$).
A faulty switch may drop a packet that meets the 
header characteristics criteria ($C_{forward}$) despite the fact that
the buffers are not full ($S_{uncongested}$).
A faulty switch may also forward a packet to an incorrect
egress interface or forward a packet that should be dropped.

The last category in Table~\ref{tab:formal_model} deals
with the interactions of the switch with
routing peers as well as with accesses from a remote
host to a management port of the switch.
A faulty switch may, for example, fail to establish
a TCP connection with a BGP peer or fail to accept
or respond to management messages from a remote host.

The high-level model presented in Table~\ref{tab:formal_model}
must be instantiated for a particular network type.
For \name, the model is instantiated for IPv4/BGP networks
(Appendix~\ref{sec:instantiation}).
For example, the forwarding characteristics and
the header transformation function are, respectively:\\
\newcommand{\ti}[1]{\textit{#1}}
\vspace{-0.1in}\\
\hspace*{0.15in}\parbox{\columnwidth}
{\small\setstretch{0.9}
    $C_{\ti{forward}} =  \ti{FIB(Header.IPv4.DstAddr)} = Q \ \land \\
     \hphantom{XX} \ti{Header.IPv4.TTL} \neq 0 \ \land \ 
     \ti{AccessControl(Header)} \neq \ti{Deny}\ \land \\
     \hphantom{XX} \ti{CheckSum(Header.IPv4)} = 0
    \vspace{0.08in}\\
   \ti{Fhdrt(Pkt(C))} = \ti{Pkt(C)}, \text{with} \
      \ti{SrcMacAddr} \xleftarrow{} \ti{EgressPortMac} \ \wedge \\
      \hphantom{XX} \ti{DstMacAddr} \xleftarrow{} \ti{NextHopMac} \
      \wedge TTL \leftarrow TTL_{in} - 1 \ \wedge \\
      \hphantom{XX} \ti{Checksum} \leftarrow
         \ti{ChecksumCompute(NewHeader.IPv4)}$
}
 \section{Primitives and Procedures}
\label{sec:prim_proc}

This section describes, at a high level,
the derivation of
diagnosis primitives and procedures for model-based diagnosis.
The next section presents the instantiation of
these primitives and procedures in \name,
our proof-of-concept implementation of a model-based
network diagnosis tool.

Diagnosis is triggered by an end-to-end
(host-to-host) failure report.
A report is generated as
a result of unexpected application behavior due
to excessive rates of
dropping or corruption of data packets by the network.
Such reports are expected to contain information
regarding the identity of source and destination hosts
and possibly other packet characteristics,
such as destination port number.

The diagnosis process must ultimately identify
the switch or link whose behavior diverges
from the expected correct behavior (\S\ref{sec:model})
and is thus the \textit{root cause} of the network failure.
The following two requirements must be met in order to enable
the detection of such divergences:

\noindent
{\bf R1)} \textit{Primitives} (mechanisms)
to collect from the operational network 
information that distinguishes faulty
functionality from fault-free functionality, as
defined in Table~\ref{tab:formal_model}.

\noindent
{\bf R2)} A way to derive from the
\textit{network model} information that is
comparable to the information collected from
the network and can thus be used to identify
which parts of the collected information is incorrect.

{\bf R1} can be met by inspecting Table~\ref{tab:formal_model}.
The first category in the table requires
primitives for collection of information
that indicates whether packets are dropped by a switch
and what are the header characteristics of those packets.
The second category requires primitives
to retrieve packet headers at ingress
and egress switch interfaces.
Categories 3-6 require primitives
to retrieve data from
the switch's data and control plane tables.
Category 7 requires primitives
to capture packets exchanged
between the switch and other entities.

The description of the required capabilities of the
primitives is a starting point for the
design and implementation of primitives for a diagnosis system.
Implementation complexity, resource requirements, and impact 
on the normal operation of the network are key
considerations for the choice of primitives.
The next section describes the primitives used by \name.

Requirement {\bf R2} can be met using
configuration analysis tools, such
as Batfish~\cite{fogel2015batfish}.
Based on switch configurations, network protocols,
and network topology, such tools can answer queries regarding
the correct network state as well as the
possible ways in which packets are forwarded and routes
are propagated.

The failure reports that trigger diagnosis contain information
(packet characteristics)
that is used to identify
the location of data packet drops.
However, the ``path'' from the root cause fault to
this location may involve a multi-step chain of \textit{symptoms},
where each symptom is a divergence from fault-free behavior.
These divergences are not necessarily packet drops.
For example, a symptom may be a switch that sends an
incorrect route advertisement despite having
a correct RIB and correct configuration.
To locate the root cause fault,
the diagnosis procedures must identify the packet drops
and then trace back through the chain of symptoms.

Packets that, according to the \textit{network} model,
should not be dropped, may be dropped by a
\textit{fault-free} switch, $S$.
This may occur if these packets are actually
not supposed to reach $S$, or due to the failure
of the rest of the network to provide to $S$
routes required to handle the packets.

To cover the scenarios above, where a fault-free switch
proactively drops packets, the diagnosis procedures 
must be able to obtain from a switch information
indicating \textit{why} packets are dropped.
Thus, the system must include primitives for
retrieving this information.
Since the switch may, in fact, be faulty, this information
cannot be trusted.
Thus, every diagnosis step that uses information from 
one switch to transfer attention to another, must
collect information from both switches
and validate both using the network model.

If only a \textit{single}
switch in the network may be faulty at any time,
potential diagnosis ambiguities can be resolved.
Since the type of faults we consider
(hardware faults and firmware or
software Heisenbugs~\cite{Gray86}) are rare,
simultaneous multiple faults are highly unlikely.
Furthermore, \S\ref{sec:evaluation} shows that
the diagnosis procedures are robust and are highly likely
to correctly handle two simultaneous faults.

In summary, the diagnosis procedure is a recursive process
that begins by locating a switch or link where
packets with the characteristics identified in the
failure report are dropped.
Next, the procedure determines whether the drops
were the correct local action.
If not, the switch is identified as faulty.
Otherwise,
the focus shifts to the switch
that forwarded the packets to the current switch,
or to a switch that provided an incorrect route or
failed to provide a required route.

 \section{\textit{N\lowercase{et}D\lowercase{x's}} Primitives \& Procedures}
\label{sec:netdx_diag}

\name is a proof-of-concept implementation of model-based
diagnosis instantiated for IPv4/BGP networks.
This section describes the instantiation of efficient
diagnosis primitives and procedures for this network type.
Implementation details are provided in \S\ref{sec:implementation}.

\begin{figure}
    \centering
    \includegraphics[width=2.4in,keepaspectratio]
                    {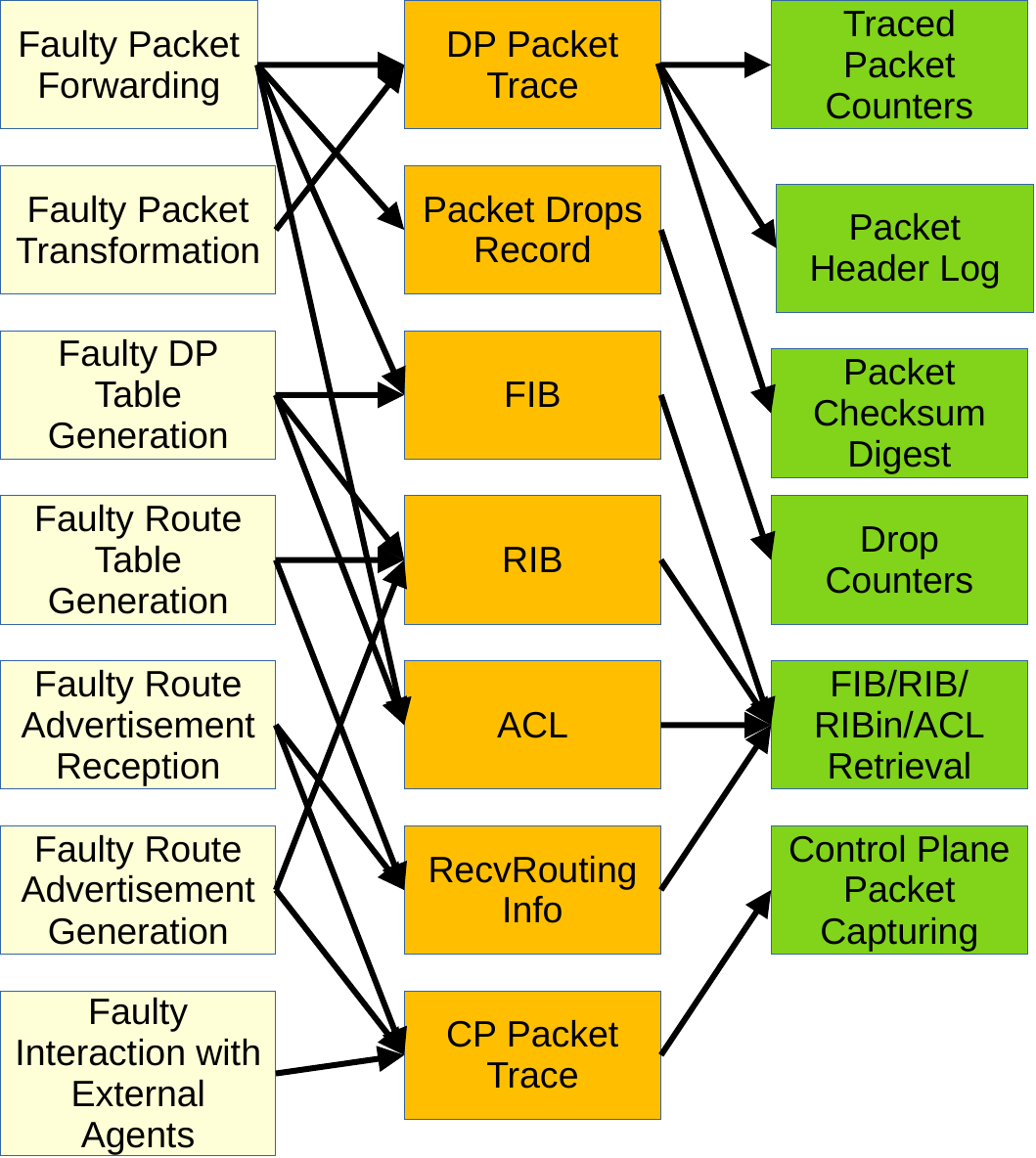}
    \caption{Derivation of \name's diagnosis primitives.
             DP: data plane. CP: control plane.}
    \label{fig:primitives}
\end{figure}

\subsection{\name's Diagnosis Primitives}
\label{subsec:dxprimitives}

Since packet forwarding and the processing of routing
information is done by the switches, the data that indicates
what is happening in the network is either stored
on the switches (e.g., tables) or passes through the
switches (e.g., data packets and route advertisements).
Thus, overhead is minimized by basing diagnosis
on data normally maintained by switches (FIB, RIB, ACL)
as well as data collected by specialized mechanisms
regarding packets that pass through the switch.
As described below, this is the basis for
the primitives used by \name.

Figure~\ref{fig:primitives} shows the derivation
of the diagnosis primitives.
The left column entries are the seven fault categories
in Table~\ref{tab:formal_model}.
The middle column entries are the sources of
information needed to diagnose particular faults.
The right column entries are the \name primitives
used to collect the different categories
of diagnosis information.

The first step of the diagnosis process is to identify
the switch or link where packets are dropped.
This can be done by comparing
counts of outgoing and incoming  packets.
Since a low rate of packet loss is generally acceptable,
simply identifying some packet loss is not sufficient.
The failure report typically specifies some
specific flow characteristics (header 5-tuple) and those
flows may account for only a small fraction of the total
network traffic.
Thus, the task is to identify the switch or link where
the rate of packet loss from the flows of interest (\textit{FoIs})
exceeds a given threshold.
Hence, there is a need to collect incoming and outgoing counts
of packets from the FoIs.

\name requires that
switches include counters at ingress and egress interfaces
that count only FoIs packets.
Hence, edge switches must be configurable
to set a \textit{trace bit}
(similar to Everflow's \textit{debug bit}~\cite{zhu2015everflow})
in
packet headers matching a given pattern.
The counters in every network switch count only packets,
henceforth \textit{traced packets}, where
the \textit{trace bit} is set.
For meaningful count comparisons, the ingress and
egress counts of every switch or link must be
from a consistent snapshot~\cite{Chandy85,Yaseen18}.
However, consistency is needed only for one
switch or one link, not the entire network
(see \S\ref{subsec:data_plane}).

Since packets may be dropped by a fault-free
switch (\S\ref{sec:prim_proc}),
once a drop location is identified,
additional information is needed to determine
whether the fault is local.
This is done based on the fact that a fault-free switch
may drop packets for several reasons:
(1)~ACL deny,
(2)~no matching forwarding entry,
(3)~the packet's TTL is 0,
(4)~incorrect IP header checksum,
and
(5)~congestion.
In all of these cases, there is an explicit decision by
a fault-free switch to drop a packet.
Hence, \name requires the switch to maintain
\textit{drop counters} that count the traced packets dropped
for each of these reasons.

\begin{table}
    \centering
    \caption{\name's diagnosis primitives.}
    \label{tab:primitives}
    \begin{tabular}{|c|l|}
    \hline
    \multicolumn{2}{|l|}{Mark packets of interest at network edge as \textit{traced packets}}  \\
    \hline
    \multirow{3}{*}{\parbox{0.36in}{ \setstretch{0.8}
       Traced \\ Packet \\ Counters}} &
	\parbox{2.6in} { \setstretch{0.8}
	Drops: no forwarding,ACL,IP header corruption, \\ 0-TTL,congestion} \\
    \cline{2-2}
    & Link silent drop detector \\
    \cline{2-2}
    & Switch silent drop detector \\
    \hline
    \multicolumn{2}{|l|}{Log of recent traced packet headers} \\
    \hline
    \multicolumn{2}{|l|}{Log of recent dropped traced packet headers} \\
    \hline
    \multicolumn{2}{|l|}{Table lookup/retrieval: ACL, FIB, RIB, RIB-in, RIB-out} \\
    \hline
    \multicolumn{2}{|l|}{Packet injection} \\
    \hline
    \multicolumn{2}{|l|}{Capture of control plane packets} \\
    \hline
    \multicolumn{2}{|l|}{Alerts for high-frequency switch state changes: FIB, RIB, connections}\\
    \hline
    \multicolumn{2}{|l|}{Alerts for resource exhaustion: ACL/FIB/RIB entries, CPU load} \\
    \hline
\end{tabular}
 \end{table}

Table~\ref{tab:primitives} presents
\name's diagnosis primitives.
In addition to the facilities described above,
they include primitives to retrieve FIB, RIB, and ACL information.
Diagnosis of some faults, such as incorrect TTL
modification, require examination of packet headers.
Hence, switches maintain locally
and provide access to
short logs of headers of recently-forwarded
and recently-dropped traced packets.
Arbitrary incorrect modifications of packets
performed by a link are detectable by the Layer~2 CRC code.
To facilitate diagnosis of arbitrary incorrect payload
modifications by switches,
switches compute and provide access to
payload checksums of specially-marked packets.
To facilitate diagnosis of incorrect route advertisements,
switches can capture and provide access to
routing packets sent to or from the control plane.
There are additional details regarding the use of these
facilities in \S\ref{subsec:switch_sw}.

\subsection{\name's Diagnosis Procedures}
\label{subsec:procedure}

\name's diagnosis procedures are derived
systematically, tracing the steps of network operation
in reverse order.
The multi-step incremental diagnosis procedure
begins with configuring switches at the network
edge or the source of the FoIs specified in
the failure report to set the trace bit for the FoIs.
It proceeds with identifying the location of
FoIs packet drops and ends with the location of
the root cause failed switch or link.

The diagnosis process is controlled by
a \textit{diagnosis manager}
running on a host in the network.
It is able to retrieve the collected
information from any switch.
The drop counters guide how diagnosis proceeds.
For example, \textit{ACL deny drops} in switch S1
of packets forwarded by switch S2
cause the diagnosis manager
to \textit{query} Batfish whether the traced packets
should ever be forwarded by S2 to S1.
A positive answer indicates that the problem
(fault or configuration error) is on S1.
A negative answer shifts the diagnosis focus to switch S2,
to determine why packets were forwarded incorrectly.
Thus, once the location of packet drops is identified,
at every point in the diagnosis process a particular
switch is the current \textit{switch under test} (SUT).
The identity of the SUT shifts to different switches
as diagnosis proceeds.

The most complex diagnosis tasks involve both the
data plane and control plane.
In these cases, the switch where the packets are dropped
has a high value of the
\textit{no matching forwarding entry} drop counter.
This triggers a sequence of steps that include
a query of whether the switch is supposed
to handle the FoIs, followed by examination of the FIB,
RIB, and RIB-in.
These steps may indicate routing information
that had been received by the switch was not processed correctly,
pointing to a fault in the switch.
However, these steps may also indicate that the expected
routing information has not been received.
In this case, this triggers steps that involve routing
peers (neighbors) that are supposed to provide the expected routing
information.

\begin{figure}
    \centering
    \includegraphics[width=3.25in,keepaspectratio]
                    {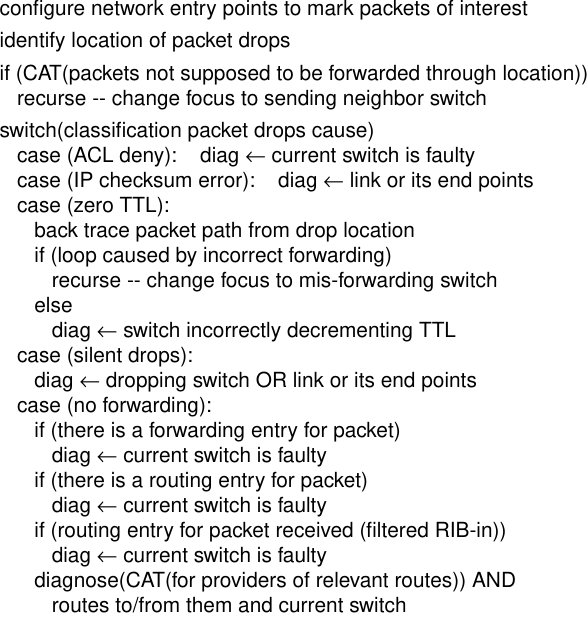}
    \caption{\name's diagnosis procedure.
             CAT: Configuration Analysis Tool.}
    \label{fig:diag_pseudo}
\end{figure}

The high-level flow of the diagnosis procedure
of \name is shown in pseudocode in Figure~\ref{fig:diag_pseudo}.

A key feature of \name is that the diagnosis
process is \textit{automated}.
This is enabled by scripts executed
by the diagnosis manager.
The scripts are triggered based on failure reports from
users, such as excessive packet drops for a particular flow.
The scripts then use the primitives iteratively to
localize the root cause to a switch or a link.
The scripts require information regarding
the topology and configuration of the network.
This information is obtained for the scripts on demand
by the diagnosis manager
from the configuration analysis tool.

The diagnosis scripts are written manually,
requiring an understanding of switch functionality,
network protocols,
and \name{}'s diagnosis primitives,
Hence, the development of scripts involves
significant human effort.
However, this is a one-time cost;
once the scripts are written, diagnosis
of network failures is automated.
For example, Appendix~\ref{appen:diag_ex}
shows the workflow of the
diagnosis procedure for the case, mentioned above,
where packets are dropped due to
\textit{no matching forwarding entry}.
This entire diagnosis procedure is
performed automatically by \name, starting with only
a failure report indicating that there are dropped packet
on some FoIs.
 \section{Implementation of \textit{N\lowercase{et}D\lowercase{x}}}
\label{sec:implementation}

\begin{figure}
    \centering
    \includegraphics[width=2.6in,keepaspectratio]{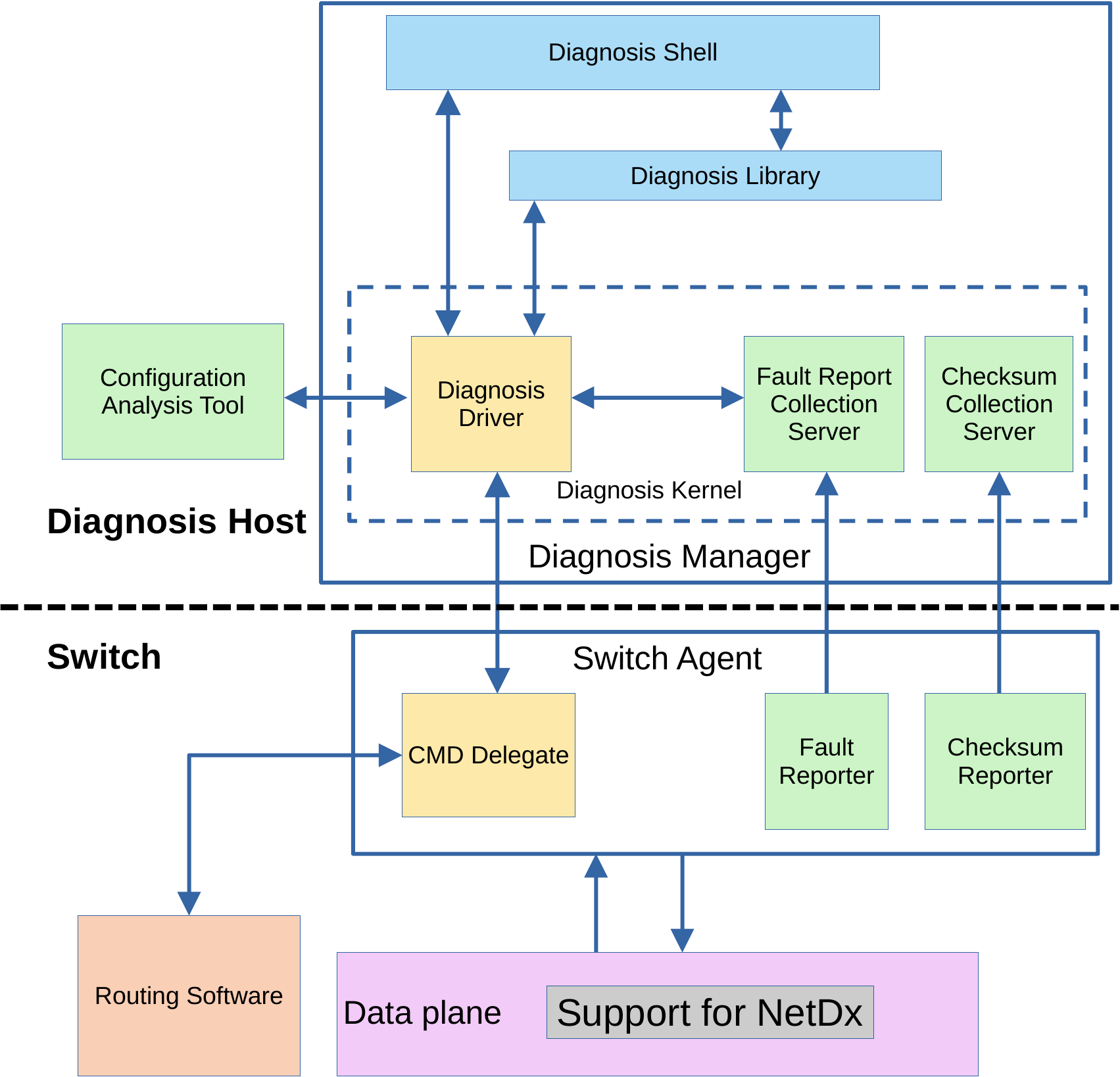}
    \caption{\name architecture.}
    \label{fig:coordination}
\end{figure}

The diagnosis procedure described in \S\ref{subsec:procedure}
requires a system that collects data on the switches
and then processes it in a coordinated manner.
This section describes an implementation of such a system.
The architecture of the system is shown
in Figure~\ref{fig:coordination}.
It includes data plane primitives implemented
on the switches, described in \S\ref{subsec:data_plane};
a \name \textit{switch agent} and related software executed
by the CPU on each switch, described in \S\ref{subsec:switch_sw};
and a \textit{diagnosis manager} executed on a host in the network,
described in \S\ref{subsec:manager}.
There is a special challenge with \name to handle
a failure of the diagnosis manager to
communicate with a switch during diagnosis;
\S\ref{subsec:switch_connection} presents our solution to
this problem.

Based on real world network faults (\S\ref{subsec:fault_set}),
we concluded that, within the framework of \name,
there is a need for
specialized detectors of anomalous switch behavior
that is otherwise difficult to detect.
These detectors operate \textit{locally} on the switches,
generating fault reports to the diagnosis manager as needed.
As described in Appendix~\ref{appen:anomaly_detectors},
they include detectors for high rate route oscillations
and for resource exhaustion.
They are part of our implementation,
but we have so far not implemented diagnosis
scripts that use their results.

\subsection{Data Plane Support for Diagnosis}
\label{subsec:data_plane}

\name requires special features of the switch data plane
that facilitate the collection of diagnosis information.
These features can be implemented in hardware.
Our prototype is designed for programmable switches
and the required functionality is implemented using
the P4\cite{p4paper, p4site} programming language.
Our implementation consists of 1432 lines of P4.
It should be noted that,
if the required data plane functionality
is implemented in some other way, switch
programmability is not necessary for \name.

\noindent
\textbf{Packet Drop Counters:}
Switches are designed to deliberately drop packets
when specific conditions related to the packet
and the switch state are met.
The switches can record such events.
Specifically, \name requires every switch to maintain,
at each ingress port, four counters that
count the \textit{traced packets} that are dropped
due to the following four reasons:
no matching forwarding (FIB) entry,
ACL deny, zero TTL, and incorrect IP header checksum.
A sliding window mechanism is used to restrict
the counts to recent packet drops.

\noindent
\textbf{Fault Report Triggers:}
The first step of the diagnosis process is to identify
the location where \textit{affected data packets} (henceforth, ADPs)
are dropped.
To minimize the latency of this step, when
\textit{traced packets} are deliberately dropped by
a switch, as described above, the switch
sends a \textit{fault report}
(\S\ref{subsec:switch_sw})
to the diagnosis manager.
This operation is triggered by the data plane.
To avoid frequent false positives, this trigger
is generated only if, during a time window,
the ratio of dropped \textit{traced packets}
to all arriving \textit{traced packets} exceeds
a given threshold.
An additional condition for the trigger is that
the number of \textit{traced packets}
received by the ingress port during the time window
must exceed another threshold.
The latter condition avoids the situation where,
for example, a fault report is triggered as a result of
a time window during which only one \textit{traced packets}
is received and it is dropped.

After a deliberate packet drop, if the conditions
for a fault report are met, the switch control plane
is notified (triggered).
This is done by forwarding a modified version
of the packet to the switch CPU.
The CPU (\S\ref{subsec:switch_sw}) sends the
fault report to the diagnosis manager.

\noindent
\textbf{Traced Packet Counters:}
Counters of \textit{traced packets}
are maintained at every switch ingress and egress port.
A sliding window mechanism is used to restrict
the counts to only recent packets.
These counters are used for silent packet drop detection
as well as for tracing the routes of FoIs.

\looseness=-1
\noindent
\textbf{Silent Drop Detection:}
Packets may be silently dropped by a switch or link.
The use of the \textit{traced packet counters}
to detect silent drops is complicated by the
fact that packets are buffered within a switch.
Thus, even if there are no packet drops, the sum
of ingress counter values is likely to be larger
than the sum of egress counter values.
Furthermore, it is not possible to read the
values of multiple counters on a switch or
the counters on two adjacent switches at
exactly the same time.

For drop detection within switches,
the problem of inconsistent counters is solved
by relying on the ability of the switch
to attach to packets meta data that is
maintained only within the switch.
Specifically, at each ingress port, the current
switch time stamp is attached to each packet.
When the packet exits the switch,
this time stamp is used to update the
\textit{virtual} clock at the egress port
through which the packet is transmitted.
Traced packet counts during two time windows
are maintained at each port.
These time windows are based on the switch clock
at the ingress ports
and the virtual clocks at the egress ports,
This allows sums of consistent counts to be computed
and used to detect silent drops within switches.

Detection of silent drops on a link is 
based on special \textit{marker packets} that are
sent by the switch agent and processed
by the data planes of adjacent switches.
Specifically, the diagnosis manager initiates
the process by sending a command to the switch agent
of one of the switches.
That switch agent sends the \textit{marker packet}
to the adjacent switch.
At the egress port of the first switch,
the current \textit{traced packet} count
is recorded in the \textit{marker packet}.
The adjacent switch ingress port intercepts the
\textit{marker packet},
records its \textit{traced packet} count
on the packet, and sends it back to the
switch agent of the first switch.
Packet drops on the link are detected
based on the two counts on the marker packet.

\noindent
\textbf{Local Packet Header Logs:}
Each switch maintains short logs of
packet headers at its ports.
At each ingress port
there is a header log of the most recent \textit{traced packets}
received at that port.
There is also a header log of the most recent
dropped \textit{traced packets} received and
deliberately dropped at that port.
For each egress port, there is a header log of the
most recent \textit{traced packets} transmitted from that port.
Each log entry contains, in addition to the packet header,
meta data such as the header length and the
reason for the packet drop.

\noindent
\textbf{Local Packet Mirroring to the Control Plane:}
To allow switch control plane software to process
select packets forwarded by the switch,
the data plane mirrors to the control plane packets
in which the DSCP field of the IP header is set to $0x14$.
As described in \S\ref{subsec:switch_sw},
this enables fast diagnosis of packet corruption.

It should be noted that the local logging of packet headers
and the very limited mirroring of packets described
above are \textit{local} to the switch.
The packet header logs are small (on the order of ten)
and are only used in rare cases during diagnosis
(diagnose drops due to TTL=0).
Local packet mirroring is used only for very few packets
(less than ten) injected only during rare diagnosis
procedures (currently only used for diagnosing packet corruption).

\subsection{The Switch Agent}
\label{subsec:switch_sw}

The switch agent is executed by the switch CPU
and provides the interface between the diagnosis
manager and the switch.
It interacts with both the switch data plane and control plane.
As shown in Figure~\ref{fig:coordination},
it consists of three modules:
the command delegate, the fault reporter,
and the checksum reporter.
The switch agent is implemented in 978 lines
of Python code.

The actions of the command delegate
are initiated by commands from the diagnosis manager.
It performs the following functions:
retrieve data plane counter values, FIB entries,
and ACL entries;
retrieve control plane RIB, RIBin, and RIBout entries;
capture routing packets sent to and from the control plane;
send information it retrieves to the diagnosis manager;
command the data plane to perform functions such as
marking packets as \textit{traced packets}
as well as performing a test for silent packet drops
on a switch or link (\S\ref{subsec:data_plane}).
The command delegate can also relay commands
from the diagnosis manager to another switch.
As discussed in \S\ref{subsec:switch_connection}, this
is used for handling situations where network failure
prevents the diagnosis manager from connecting to some switches.

The fault reporter receives packet drop notifications
from the data plane (\S\ref{subsec:data_plane})
and relays them to the diagnosis manager.
To avoid fault report storms,
after sending a fault report, the fault reporter sets a flag in the
data plane that disables further drop notifications.
The diagnosis manager eventually unsets this flag via
the command delegate.

The checksum reporter is used for diagnosing corruption
of packet payloads.
This diagnosis involves injecting packets with
the DSCP field set to $0x14$, which are 
then mirrored at every switch to the control
plane (\S\ref{subsec:data_plane}).
The checksum reporter receives these packets
and sends a limited number of checksums to the diagnosis manager.
The diagnosis manager uses these checksums
to identify the corruption location.

Once a failure report is received, diagnosis
requires packets matching the FoIs' characteristics (5-tuple)
to be continuously transmitted.
However, the packets' source may be remote
and under different administrative control.
Hence, there is need to be able to inject
at a data center edge switch packets that match the ADPs.
This injection is performed by the switch agent,
triggered by a command from the diagnosis manager.

\subsection{The Diagnosis Manager}
\label{subsec:manager}

The diagnosis manager is executed on a \textit{diagnosis host},
where it coordinates the diagnosis process and
interacts with the human network operator.
The diagnosis manager also interacts with the
Batfish~\cite{fogel2015batfish} configuration analysis tool,
from which it obtains responses to queries
about what the switches should be doing.
The diagnosis manager is implemented in 1688 lines
of Python code, while the diagnosis scripts (see below)
consist of 1036 lines of Python code.

The diagnosis manager provides an API that allows
sending commands to switch agents, receiving
information retrieved by switch agents,
and interaction with Batfish.
Diagnosis is generally performed by scripts
that use this API.
A command line interface (CLI) allows direct
invocations of primitives and scripts as well
as the instantiation and invocation of new scripts.
Multiple scripts can be stored as a \textit{script library}.
Scripts can invoke other scripts.
All the diagnoses reported in this paper were
performed by such a \textit{diagnosis library}.
This library includes a top-level script that
is invoked with a description of the failure report
and returns a diagnosis.

The fault injection campaign (\S\ref{subsec:inj_campaign})
showed that, in a few outlier cases, \name failed to produce
a diagnosis if the network routing state 
was in the process of changing.
With the \name scripting mechanism this was easy to fix.
Specifically, we modified the top level script to run each diagnosis
at least twice with a short delay between runs.
The diagnosis process completes only when 
two consecutive runs yield the same result.
Since a diagnosis run typically completes in seconds,
the additional latency is not significant.
After this modification, all
the outlier cases resulted in correct diagnosis results.

\subsection{Handling Disconnected Switches}
\label{subsec:switch_connection}

During diagnosis, the diagnosis manager (\textbf{DM})
typically connects to the switch agents on multiple switches.
However, network failure may prevent the connection
to some of the switches,
thus potentially blocking the diagnosis from proceeding normally.
To avoid such blocking,
the original diagnosis is transformed to
a diagnosis of the connection failure between the
diagnosis host (\textbf{DH}) and the disconnected switch (\textbf{DS}).
This failure may indicate
that \textbf{DS} is faulty but may also be a result
of faults on other switches.
Thus, as described below, \name includes a special procedure
to handle this scenario.

\textbf{DH} is assigned two IP addresses:
the \textit{primary} (\textit{PIP}), that is used normally,
and the \textit{secondary} (\textit{SIP}), that is used
as described below.
Throughout the network, the configuration of 
routing to these two IP addresses is normally identical.
The procedure is based on forming a
\textit{statically-routed path} (henceforth, \textit{SRP}),
composed of a sequence of static routes, between
\textbf{DH} and \textbf{DS}.
This is done using \textbf{DH}'s \textit{SIP}
and \textit{SIP}s assigned to the switches as needed by \textbf{DM}.

Upon failure of \textbf{DM} to connect
to a switch, it initiates diagnosis of this failure.
This is done hop-by-hop, starting from the switch closest
to \textbf{DH}, along a path acquired by querying Batfish,
using \textit{PIP}s.
At each hop, static routes are added between \textbf{DH}'s
SIP and the SIP that is to be assigned to \textbf{DS}.
This phase of the procedure identifies the disconnected switch
closest to \textbf{DH} along this route.
If this switch is not the original \textbf{DS},
the procedure is recursively invoked
with the switch closer to \textbf{DH} now being the focus.
We denote by \textbf{N} the physical neighbor switch of \textbf{DS},
which is closer to \textbf{DH} on the route
from \textbf{DH} to \textbf{DS}.

Since \textbf{DS} and \textbf{N} are physical neighbors, they
can communicate regardless of routing,
unless one of them or the link between them is faulty.
Based on the procedure described above, \textbf{DM}
can connect to \textbf{N} using their \textit{PIP}s.
The command delegate on \textbf{N} is used to relay
commands from \textbf{DM} to \textbf{DS}.
If \textbf{N} is unable to connect to \textbf{DS},
this locates the fault to
\textbf{DS}, \textbf{N}, or the link between them.
Otherwise, using this mechanism, \textbf{DM} attempts to cause
\textbf{DS} to install a static route from itself
to \textbf{DH}'s \textit{SIP} via \textbf{N}.
This ensures that packets from \textbf{DS}
to \textbf{DH} are not routed on a path
that may include a faulty switch.
As a result, \textbf{DH}, using its \textit{SIP},
is able to communicate with \textbf{DS}.

After the steps above,
\textbf{DM} initiates the diagnosis
of a connection failure from
\textbf{DH} to \textbf{DS}'s \textit{PIP}
and then from
\textbf{DS} to \textbf{DH}'s \textit{PIP}.
For this diagnosis, \textbf{DM}
uses \textbf{DH}'s \textit{SIP}
to communicate with the switch agent on \textbf{DS}.

 \section{The \textit{P\lowercase{odnet}} Emulation Platform}
\label{sec:emulation}

The implementation of \name required
a network with data and control planes
that are as close as possible to those of a real network.
It also required the ability to easily
modify the data and control planes as well as to
later be able to test the operation of \name
under a variety of faults.
These requirements mandated the use of a simulated
or emulated network.
To facilitate data plane modifications,
switches with programmable data planes~\cite{p4paper}
were the obvious choice.
For a realistic control plane, it was desirable to
use routing software that is also used in real networks.
As in real networks, an independent instance
of this software must run on each switch.

Based on the above considerations, available
existing tools~\cite{lantz2010mininet, ns3, fan2017ns4, bai2018ns4}
were not suitable for our purpose.
This led to the development of \emuname,
an emulator of a network of switches with data planes
that are programmable using P4~\cite{p4paper}
and a control plane based on FRR~\cite{frr}.
Since both P4 and FRR are used in real networks
and \name is operational on \emuname,
it is expected that our \name implementation
can be ported to a real network with limited effort.
Appendix~\ref{appen:podnet} presents
additional details regarding \emuname.
 \section{\name Diagnosis Example}
\label{sec:example}

\begin{figure}
   \centering
   \includegraphics[width=\columnwidth,keepaspectratio]
   {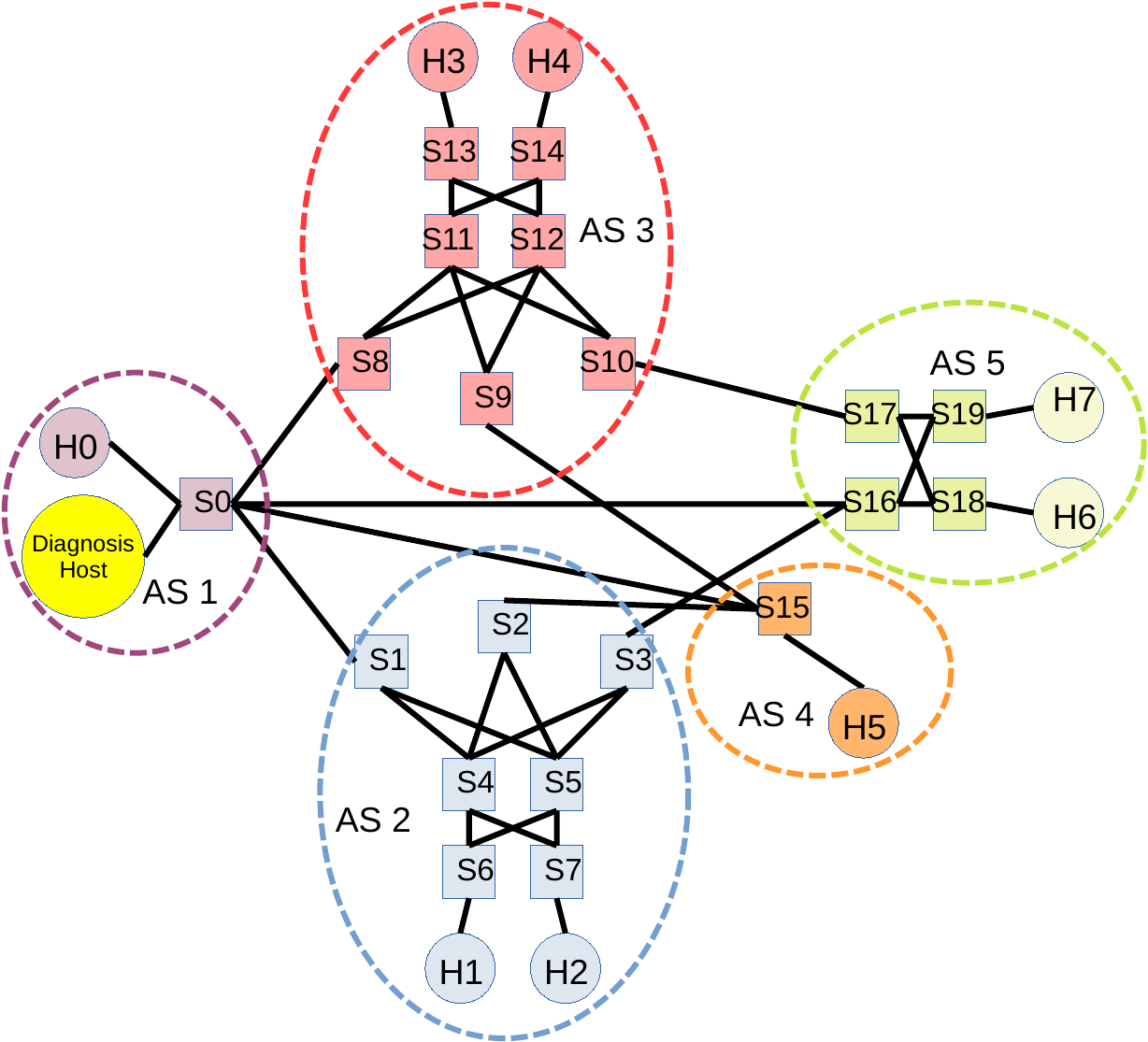}
   \caption{Topology of the network used in \S\ref{sec:example}
            and \S\ref{sec:evaluation}.}
   \label{fig:topology}
\end{figure}

To further explain the operation of \name,
this section describes the diagnosis by \name of
an example fault.
We use a network
whose topology is shown in Figure~\ref{fig:topology},
emulated on \emuname (\S\ref{sec:emulation}).

The fault is that the IGP and BGP routing daemons of
switch S10 (AS-3) crash (fault class~7 in
Table~\ref{tab:fault_dataset}).
As a result, switch S17 (AS-5) withdraws the routes to AS-3 and AS-4
from the switches in AS-5.
AS-3 and AS-4 switches still have routes to AS-5
through AS-2.
However, due to policy configurations, S0 (AS-1) and S3 (AS-2) do not
advertise to AS-5 routes to ASes other than themselves.
Hence, AS-5 switches do not have any routes to AS-3 and AS-4.
Thus, communication between
hosts H7/H6 (AS-5) and hosts H3/H4/H5 is disrupted.

We injected the fault above
by killing the routing daemons on switch S10.
As a result, a user-level failure report indicated
that host H7 cannot reach host H3.
In response,
\name's top-level diagnosis script was invoked,
specifying the source and destination hosts.
The diagnosis script configured switch S19 (AS-5) to set the trace
bit of packets from H7 to H3.
On switch S19, the \textit{no forwarding entry drop counter}
value exceeded the threshold and the
switch sent a fault report to the diagnosis manager (\textbf{DM}).
As a result, the script on \textbf{DM}
invoked the \textit{diagnose no forwarding drop} script,
targeting switch S19.
The script retrieved routing information from switch S19
and determined that the required route was missing.
The script then queried Batfish
for the switches that were supposed to propagate the route to S19 and
the response pointed to switch S17 (AS-3).

Based on the above, the script invoked the
\textit{diagnose bgp route adv missing} script, targeting
switch S17.
This script retrieved routing information from S17 and found
that S17 did not have the route.
By querying Batfish, the script found that S10 (AS-3)
should provide the route to S17.
The script retrieved status information regarding S10 from S17
and found that the BGP
connection between S17 and S10 was down.
This caused the invocation of the
\textit{diagnose bgp routing neighbor down} script targeting
S17 and S10.
The script attempted to establish a TCP connection with S10.
Since the routing daemon on S10 had crashed, the rest of
the network lost routes to S10 and this attempt failed.

Due to the failure to connect to S10,
the diagnosis script invoked \textit{diagnose connectivity
loss of switch} script (\S\ref{subsec:switch_connection}),
with S10 as the ``disconnected switch'' (\textbf{DS}).
As described in \S\ref{subsec:switch_connection},
this ended up with the invocation of the script
to diagnose packet drops between the diagnosis host
(\textbf{DH}) and S10.
This involved configuring S0 (AS-1) to set the trace bit
packets destined to S10.
After a short delay, \textbf{DM} received from S0 a
\textit{no forwarding entry} packet drop fault report.
This led to a new invocation of the
\textit{diagnose no forwarding drop} script, targeting S0.
This script ultimately determined that
S10 should provide a route for itself to its neighbors while it
does not.
This led to the script attempting and failing to
retrieve routing information from S10,
identifying S10 as the root cause.
 \section{Evaluation}
\label{sec:evaluation}

This section presents preliminary evaluation and validation
of \name.
It involved injection of faults using \emuname
and the execution of the diagnosis script by \name.
The experiments used an emulated network,
whose topology is shown in Figure~\ref{fig:topology},
consisting of five autonomous systems,
twenty switches, and nine hosts.
We collected a set of faults that were
detected in a production network of a major cloud service provider.
The fault set and the a summary of the diagnosis
coverage of \name is presented in \S\ref{subsec:fault_set}.
The fault used as an example in \S\ref{sec:example} is based on
one of the faults in this fault set.
In practice, network diagnosis is often done manually
by network operators, using tools such as ping and traceroute.
A comparison of the diagnosis of the fault in
\S\ref{sec:example} with \name vs. manual
diagnosis is presented in \S\ref{subsec:manual_diag}.
An automated fault injection campaign and its results
are presented in \S\ref{subsec:inj_campaign}.

\begin{table}
	\centering
	\captionsetup{justification=centering}
	\caption{A fault set collected from a major cloud service
	         provider.}
	\label{tab:fault_dataset}
	{\setlength{\tabcolsep}{3pt}
\hyphenpenalty=10000
\raggedright
\renewcommand{\tabularxcolumn}[1]{m{#1}}
\begin{tabularx}{\columnwidth}
{@{} d{1}@{}|@{} d{2}@{}| >{\raggedright\arraybackslash}X }
\hline
\multicolumn{1}{@{}c@{}|}{
   \begin{tabular}{c}\textbf{Fault} \\ \textbf{Class}\end{tabular}}  &
   \multicolumn{1}{|@{}c@{}|}{
   \begin{tabular}{c}\textbf{\# of} \\ \textbf{Faults}\end{tabular}}  &
   \multicolumn{1}{|c}{\textbf{Description}} \\
\hline
1 & 18 & A switch failed and dropped packets. \\
\hline
2 & 2 & A link failed and droped packets.  \\
\hline
3 & 6 & A switch or a link corrupted packets. \\
\hline
4 & 1 & Software fault caused installation of ACL rules to fail. \\
\hline
5 & 1 & Routing software crash caused routes missing. \\
\hline
6 & 1 & BGP route table was inconsistent because of software fault. \\
\hline
7 & 1 & FIB entry corrupted by data race of software. \\
\hline
8 & 3 & Route oscillation caused intermittent route missing. \\
\hline
9 & 6 & Faults that require primitives for diagnosing congestion. \\
\hline
10 & 3 & Faults that need emulation for overlay network. \\
\hline
11 & 1 & A hardware fault that caused extremely high delay on a switch rather than packet drops. \\
\hline
12 & 1 & The effect of the fault was transient. \\
\hline
13 & 3 & Route filter configuration errors caused routes missing. \\
\hline
14 & 1 & An incorrect configuration change caused BGP connections down. \\
\hline
15 & 4 & Packets were denied by ACL due to ACL configuration errors. \\
\hline
\end{tabularx}
}
 \end{table}

\subsection{Collected Fault Set}
\label{subsec:fault_set}

\looseness=-1
Table~\ref{tab:fault_dataset} presents
a set of 52 faults that were detected and diagnosed
in the enterprise network
of a major cloud service provider over multiple years.
These faults are grouped into 15 equivalence classes.
The collected data included 16 faults, excluded
from Table~\ref{tab:fault_dataset},
that were automatically detected and corrected
by built-in mechanisms on the switches and thus did
not require external diagnosis.
The faults in classes 1, 2, 3, and 11 involved only
the data plane.
The diagnosis of the faults in classes 4-10 and 13-15
also involved the control plane.

Out of the 52 faults in Table~\ref{tab:fault_dataset},
19 faults are currently not within the scope of
\name, as defined in \S\ref{sec:intro}.
These include faults in classes 10, 11, 12,
which, respectively, involve overlay networks,
cause packet delays rather than packet drops,
or are transient faults.
In addition, the six faults in class~9 are
also not within the current scope of \name, since
\name does not have the mechanisms for diagnosing
the root causes of congestion.
The eight faults in class 13, 14, 15 were configuration
errors, which are handled by network verification
tools~\cite{beckett2017minesweeper,
fogel2015batfish, HSA, ERA, NoD, ye2020hoyan, tian2019jinjing}
and are thus not targets for a tool focused on diagnosing
hardware and software faults in operational networks.

There are 33 faults that are potentially within
\name's domain in equivalence classes 1-8.
For each class, we injected a fault
in our emulated network that reproduces the
effect of the fault on the switches and links.
In each case, the input to \name corresponded
to the failure report that a user would provide
to the network operators.
Specifically, this input identifies
the affected flows by providing source, destination,
and possibly other packet characteristics.
In all eight cases, the diagnosis script executed by \name
identified the fault that was root cause of the failure.
However, three out of the 33 faults were intermittent
and there were no records regarding their frequency.
Hence, we could not determine whether they
could be diagnosed by \name.
To be conservative in our reporting, we count them
(3 out of 33) as diagnosis failures.

\subsection{Comparison with Manual Diagnosis}
\label{subsec:manual_diag}

We used the example described in \S\ref{sec:example}
to compare \name with manual diagnosis.
We asked three Computer Science Ph.D. students whose research
focuses on computer networks and are unaffiliated with our
group to act as three independent operators to perform the
manual diagnosis.
We informed them of the end-to-end failure report above.
They were allowed to use ping and traceroute
on any host or switch to probe any destination and they also
had access to all switches.
We also ran \name's diagnosis script that produced the
diagnosis automatically.

For comparison, we measured the time to reach diagnosis
and the number of commands invoked:
Person 1: 33:48 minutes, 38 commands;
Person 2: 26:32 minutes, 32 commands;
Person 3: 14:44 minutes, 18 commands;
\name: 1:38 minutes, 77 commands (primitives).

As expected, manual diagnosis was much longer.
Moreover, this experimental setup artificially
increases \name's diagnosis time since \emuname,
where every switch is a complex simulator, is extremely slow.
In a real enterprise network, where packet latency
is, at most, a few milliseconds, \name's diagnosis time
would certainly be less than five seconds.
There are also many factors that impact the manual
side of the experiment.
For example, while our testers are knowledgeable
regarding networks, they are not experienced network operators.
However, it should be noted that the fault used
in this experiment is based on a real fault that happened
in the global WAN of a major cloud service provider.
It took them more
than 30 minutes to identify the faulty switch.

\subsection{Fault Injection Campaign}
\label{subsec:inj_campaign}

To test the coverage and robustness of \name,
we developed an automated fault injection
campaign as described below.
In the first part of the campign, only a single fault
at a time was injected.
While it may appear that nothing can be gained
by injecting faults from the fault model
used to derive the diagnosis procedures,
this was invaluable for
enhancing some of the diagnosis procedures
(\S\ref{subsec:manager}),
identifying and correcting bugs in our implementation,
and estimating
the latency of diagnosis (\S\ref{subsec:campaign_results}).
In the second part of the campign two simultaneous faults
were injected.
Since \name was developed to handle only a single fault at
a time, the goal was to check its ability to handle
more complex scenarios.

\subsubsection{Experimental Setup}

We built a test harness on \emuname (\S\ref{sec:emulation})
that injects faults in a network and evaluates
the \name diagnosis outcomes for these faults.
The harness is based on a setup similar to
Pingmesh\cite{guo2015pingmesh}
consisting of processes executing on all the hosts,
sending ICMP probes to all other hosts.
If one of these processes detects multiple consecutive
dropped probes, it sends a failure report
to an \textit{injection campaign controller}
running on the diagnosis host.
Each pingmesh process is limited to sending
at most one failure report until it is reset.
The communication between the pingmesh processes and
the campaign controller is out of band, so that it
is not affected by the fault injection.
The campaign controller can cause the injection of
ten different fault types on any of the switches.

It is important to note that Pingmesh and the
\textit{injection campaign controller} are
not part of the diagnosis process and are
\textbf{not part of \name}.
In this setup Pingmesh is used to represent a user
who reports an end-to-end (host-to-host) network failure.

An injection run begins with the campaign controller
using the facilities of \emuname to inject
one or two faults in switches.
The controller then waits a specified amount
of time to receive failure reports
from the pingmesh processes.
If multiple failure reports are received,
one of them is selected at random
to invoke the top-level \name diagnosis script.
When a diagnosis result is produced by \name, it is compared
to the known fault injection locations to determine
whether the diagnosis is correct.
The fault is then reversed (undone) and the pingmesh is reset.
If two faults were injected, a new diagnosis run is initiated
to diagnose the second fault.
Otherwise, a new injection run is started.

\begin{table}
	\centering
	\vspace{-1pt}
	\caption{Fault injection campaign results.}
	\label{tab:fault_injection_campaign}
	{\setlength{\tabcolsep}{3pt}
\setstretch{1.1}
\begin{tabular}
{| l | d{3} | d{1} |}
   \hline
   \multicolumn{1}{|@{}c@{}|}{Fault Type} &
   \multicolumn{1}{c|}{\#primitives} &
   \multicolumn{1}{c|}{\parbox{\widthof{reports}}
      {\setstretch{0.8}\vspace{2pt}\#fault \\ reports}} \\
\hlineB{2.5}
   SilentDropInSwitch & 27 & 0 \\
   \hline
   SilentDropOnLink & 22.3 & 0 \\
   \hline
   CorruptionOnLink IP & 142.0 & 10 \\
   \hline
   IncorrectDecrementTTL & 27.6 & 10 \\
   \hline
   PacketPayloadCorruptionInSwitch & 22.3 & 0 \\
   \hline
   IncorrectForwardingDrop & 33.2 & 10 \\
   \hline
   FIBDiscrepancy & 24.1 & 8 \\
   \hline
   IngressBgpUpdateModification & 90.5 & 10\\
   \hline
   BgpNeighborMissing & 131.5 & 8\\
   \hline
   EgressBgpUpdateModification & 108.1 & 10\\
   \hline
\end{tabular}
}
 \end{table}

\subsubsection{Fault Injection Campaign Results}
\label{subsec:campaign_results}

Table~\ref{tab:fault_injection_campaign} shows
ten fault types
that cover all the categories
in the fault model.
In the single fault injection campaign, faults from each 
fault type were injected in a random location
until there were ten injections that resulted in
failure reports from pingmesh.
In the double fault injection campaign,
there are 100 injection runs, with two faults injected per run.
The types and locations of the two faults
are picked at random.
In both campaigns all the faults were correctly
diagnosed by \name.

\noindent
\textbf{Single fault campaign.}
Table~\ref{tab:fault_injection_campaign} shows the
results from the single fault injection campaign.
The table shows, for each fault type,
the average number of
primitives that the diagnosis manager invoked
in switch agents to diagnose the fault, as well as
the number of diagnoses in
which the switch where ADPs were dropped was
identified by a fault report from that switch.

The results show that in most cases the switch
where ADPs were dropped was identified by a fault
report from that switch.
This shows the value of the proactive fault reporting
mechanism that eliminates the need for the diagnosis
manager to retrieve information from many switches
just to find the switch dropping the ADPs.
Out of the 100 diagnoses, 6 involved dealing with
a disconnected switch (\S\ref{subsec:switch_connection}),
demonstrating the need for this mechanism.
The ``\#primitives'' column provides information
regarding the expected diagnosis latency on a real network.
Specifically, each primitive translates to a latency
of a network round trip plus the processing time
for the primitive on the switch.
\textit{Thus, diagnosis latency with a real network
will be on the order of, \textbf{at most, seconds}}.

\noindent
\textbf{Double fault campaign.}
In the double fault injection campaign, the diagnosis
of each fault pair involved two invocations of \name.
The first invocation identifies one of the faulty switches,
that switch is then ``repaired,'' and a second
invocation identifies the second faulty switch.
A key point is that in first invocation the existence
of two faults did not prevent a correct diagnosis result.
However, the average number of primitives required
for the first invocation was 98.01, while it was
only 36.26 for the second.
This indicates that the diagnosis of the first fault
was a more complex process than that of the second.
For example, 11\% of the first fault diagnoses required
dealing with a disconnected switch, but this
was required for only 1.8\% of the second fault diagnoses.
The proactive fault report mechanism was utilized
in 71 out of the 100 double fault injection runs.
 \section{Discussion}
\label{sec:discussion}

This section summarizes some of the key results
and limitations of this work.

\noindent
\textbf{Scalability of \name:}
Scalability is a key requirement for a network diagnosis mechanism.
\mbnd enables scalability by
guiding targeted collection of relevant data,
avoiding the need to collect information from the entire network.
Two key mechanisms of \name further enhance scalability.
(I)~The \textit{trace bit} mechanism
(\S\ref{subsec:dxprimitives},
based on Everflow's \textit{debug bit}~\cite{zhu2015everflow})
eliminates the need to configure every switch
to target the flows of interest.
(II)~The fault reporting mechanisms in the switches
(\S\ref{subsec:data_plane}, \S\ref{subsec:switch_sw})
allow switches with relevant information to proactively
send it to the diagnosis manager, eliminating the
need for the diagnosis manager to collect information
from every switch that \textit{potentially} has relevant information.

\noindent
\textbf{Overhead of \name:}
A key to the low overhead of \name is that it avoids
packet mirroring and logging to servers
as well as piggybacking diagnostic information on normal packets.
Only a small amount of data is collected locally on each switch.
During diagnosis, only minimal interactions are
required between the diagnosis manager and switch agents,
involving negligible network throughput.

\noindent
\textbf{The cost of developing diagnosis scripts:}
A limitation of this work is that human experts are
required to develop diagnosis scripts.
However, as explained earlier (\S\ref{sec:prelim}),
\mbnd enables and guides this development,
thus greatly reducing the required human effort.
Furthermore, these scripts are developed once and can then be used
numerous times
in multiple locations, without experts.
Importantly, the diagnosis procedures depend only on
network protocols, network topology,
and switch configurations.
They do not depend on the detailed switch implementation.
Thus, human experts are required only to support
additional protocols, not for every new switch implementation.

\noindent
\textbf{Switch hardware requirements:}
A limitation of \name is that it requires
switches capable of operations
(\S\ref{subsec:dxprimitives}, \S\ref{subsec:data_plane})
that are not
all available on many current switches.
However, as explained earlier,
while \name is an efficient low overhead implementation of \mbnd,
\mbnd can be implemented
using different mechanisms for data collection.
Furthermore, the data collection requirements
of \mbnd can guide future switch developments,
suggesting primitives that should be added to ASICs used
to implement these switches.
Finally, as demonstrated by our prototype,
the ability to support a new generation of network diagnosis
mechanisms may be a motivation to
consider programmable switches.

\noindent
\textbf{Evaluation:}
A limitation of this work is that we were not able to deploy
and validate \name on a real network.
However, our prototype implementation does show
that an efficient implementation of \mbnd is feasible
on existing hardware (P4 switches).
The extensive fault injection campaign (\S\ref{subsec:inj_campaign})
demonstrates the robustness of the implementation.
While the comparison with manual diagnosis
(\S\ref{subsec:manual_diag}) did not
match up \name with highly-experienced network operator,
such operators are often not immediately available
in real-world deployments.
A key advantage of \mbnd and \name is that they enable
automated diagnosis without such experts.

 \section{Related Work}
\label{sec:related}

\noindent
\textbf{New Data-plane Diagnostic Primitives:}
Some prior works~\cite{flowradar, narayana2017marple, gupta2018sonata}
propose data plane primitives 
implemented on the switches that enable
new queries that allow network operators to inspect
subtle network behaviors.
Compared to \name,
they can collect finer-grained
information at the cost of
higher overhead on the switches.
They are usually used to analyze
network performance issues rather than network failures.
Other mechanisms~\cite{INT, PINT, deepInsight, jeyakumar2014minions,
tammana2016pathdump, tammana2018switchpointer}
piggyback diagnostic information on packets and require
the destination hosts to send the information to collection servers.
This involves overhead on the hosts and collection servers,
increased network bandwidth use, and potentially
require packet fragmentation.
None of these mechanism can diagnose control plane faults and the
interaction between data and control planes.

\noindent
\textbf{Data Plane Log Analysis and Monitoring:}
Similarly to Cisco's NetFlow~\cite{xNetFlow},
these mechanisms log
data plane information from switches to
collection servers for later diagnosis.
In some cases, switches send packet-level information
to the collection servers~\cite{zhu2015everflow, handigol2014netsight,
handigol2012ndb, fonseca2007xtrace}.
Spidermon\cite{wang2022spidermon} optimizes the collection
of logs using wait-for relations for specific performance faults.
dShark~\cite{dShark} proposes a data processing
engine that facilitates the log analysis.
In general, these mechanisms incur high overhead on the switches,
networks, and collection servers.
Monitoring systems, such as
Pingmesh\cite{guo2015pingmesh} and NetBouncer~\cite{NetBouncer},
use servers to send probes across the network to triangulate the
location of data plane faults.
Dapper~\cite{ghasemi2017dapper}
and Trumpet~\cite{moshref2016trumpet} also use host-based
monitoring but leverage inference algorithms to reconstruct
the network state.
They are helpful for detecting the symptoms
of faults but do not provide root cause diagnosis.
Most log analysis and monitoring
systems do not focus on control plane faults or the interaction
between the data and control planes.

\noindent
\textbf{Route Log Analysis:}
Some works propose tools that analyze routing dynamics in
the internet by processing and visualizing logs of
BGP updates~\cite{lad2004linkrank, colitti2005bgplay,
blazakis2006bgp-inspect}.
These are particularly useful for diagnosing the global WAN,
where no single entity can access all parts of the network.
They incur high overhead
and do not cover data plane faults
or the interaction between data and control plane faults.

\noindent
\textbf{LLM-based Diagnosis:}
Some recent works propose the use of Large Language Models to
automate diagnosis for cloud systems.
NetAssistant\cite{haopei2024assistant} uses
LLM to automate the interaction between users and the diagnosis
system.
Since the actual diagnosis is done by proprietary
in-house workflows, we cannot compare it with \name.
RCAcopilot\cite{chen2024rcacopilot} only shows the compound F-1
score for diagnosing multi-class distributed systems.
It is unclear how effective it is in diagnosing computer networks.

\noindent
\textbf{Summary:}
\name is advantageous over prior work in providing
automation, low overhead, and coverage of
data plane and distributed control plane interactions.
Further, \name represents a new top-down model-based paradigm
that is completely different from past work which mostly
consists of bottom up debugging primitives that must be combined
manually by experts to do end to end root cause diagnosis.
 \section{Conclusion}
\label{sec:concl}

This paper introduces
\textit{model-based network diagnosis},
a foundation for tools that
\textit{\textbf{automatically}} identify
the root cause of disruptions in operational networks due to
switch hardware and software faults.
Diagnosis is based on identifying deviations
from a formal end-to-end model of packet
forwarding and routing.
This model is based on network protocols,
network topology, and switch configurations.
The model is used to derive a
functional fault model
for switches that describes how switch functionality can be disrupted.
This approach is enabled by
\textit{leveraging configuration analysis tools}
(e.g., Batfish)
that provide a representation of the network model
that is directly comparable to what is observable
on the operational network.

As an example of \textit{model-based network diagnosis},
we implemented \name, a system
that takes a high-level symptom of a network failure as input
and
automatically pinpoints
the root cause failed switch or link.
\name's diagnosis primitives and procedures are derived
systematically from the network model.
The novel basis of \name allows it to diagnose
networks with distributed control planes, tracing
through interactions between the data and control planes,
to identify the root cause of network failures.
\name's choice of mechanisms show a way to \textit{instantiate}
model-based network diagnosis with essentially
no performance overhead.

Unlike earlier
approaches that use ideas from streaming databases~\cite{dShark}
or probabilistic inference~\cite{NetBouncer}, \name uses
ideas from the dependability community, such as explicit fault
models, minimizing intrusion, and fault injection
campaigns~\cite{Chang70, Thatte78, rio, bwj_ft_book, Arlat03}.
Unlike classical work in the dependability community, however,
\name uses the fact that networks are simple enough computing
artifacts to derive what their correct behavior should be.

The evaluation of \name on a real world fault dataset shows that
it is able to diagnose most network failures.
An automated fault injection campaign was invaluable
in the development of \name and demonstrates
the robustness and coverage of \name.
It also shows that, on a real network, diagnosis
can be performed \textit{\textbf{in seconds instead of hours}}.
Future work on \name includes extending the model-based paradigm
to overlay networks, other IP features, and other networks.
 \appendices

\vspace{0.1in}

\section{No Forwarding Diagnosis Example}
\label{appen:diag_ex}

\begin{figure*}[h]
        \centering
        \includegraphics[width=2\columnwidth,height=3.0in,
	                 keepaspectratio]
	                {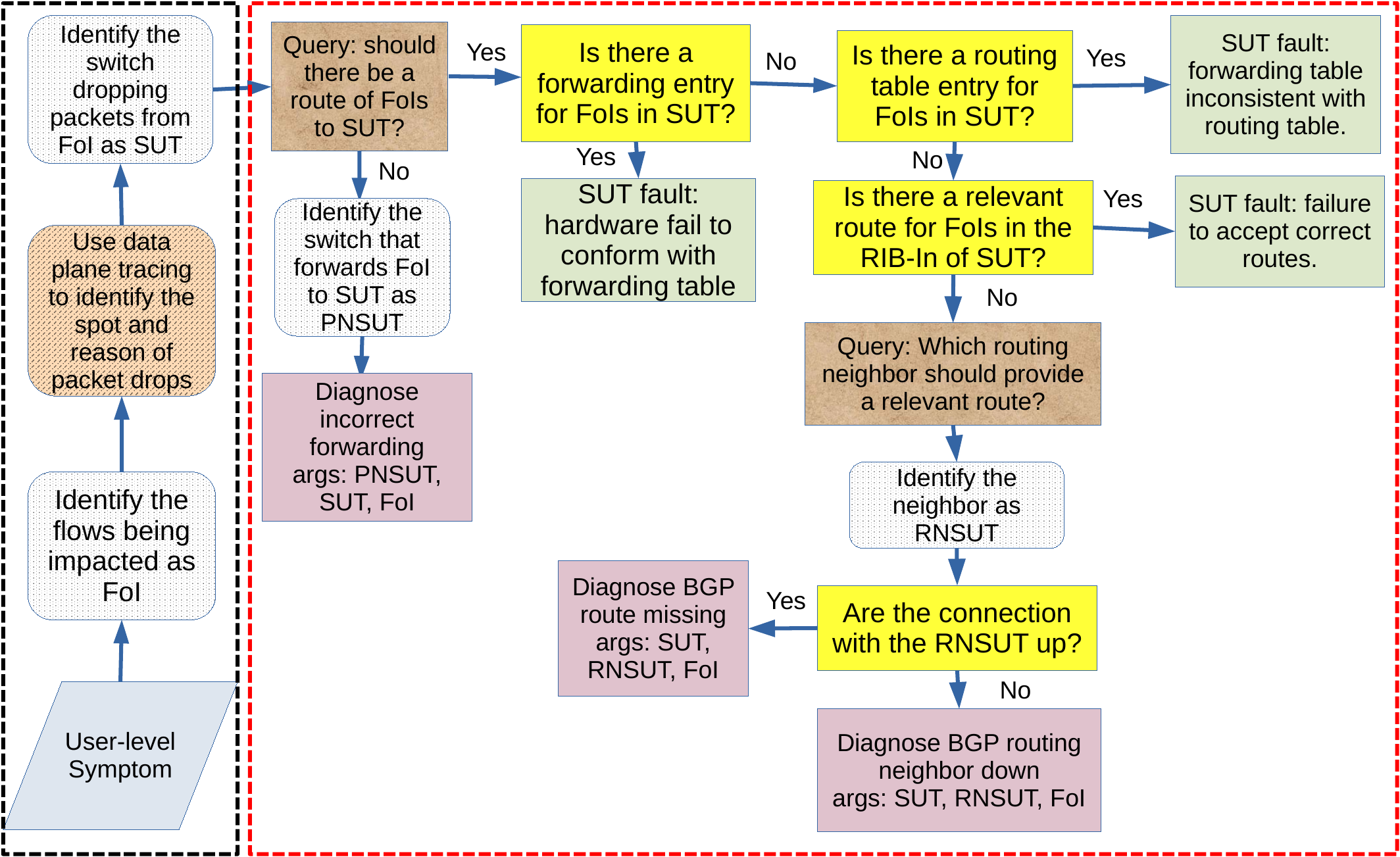}
        \caption{Diagnosis of packet drops due to no matching
	   forwarding entry. SUT: Switch Under Test;
	   FoI: Flow of Interest; PNSUT: Physical Neighbor of
	   the SUT; RNSUT: Routing Neighbor of the SUT.}
        \label{fig:no_forwarding}
        \vspace{-1em}
\end{figure*}

Figure~\ref{fig:no_forwarding} shows the workflow of the
diagnosis procedure for the \textit{no forwarding} case.
\name's general diagnosis procedure is applicable to
all routing protocols.
However, diagnosis of the control plane involves
capturing and decoding routing messages.
The procedure shown in Figure~\ref{fig:no_forwarding}
is specialized to BGP.
In one outcome of the procedure in Figure~\ref{fig:no_forwarding},
the conclusion is that there is a broken connection
with a routing neighbor.
This is not the end of the diagnosis procedure.
Rather, this triggers an invocation of a procedure to
diagnose connectivity between the two routing neighbors.
This is a recursive invocation of the same procedure
that is initially triggered by the original failure report.

 \section{Specialized Anomaly Detectors}
\label{appen:anomaly_detectors}

As mentioned in \S\ref{sec:implementation},
there is a need to implement
specialized detectors of anomalous switch behavior
that is otherwise difficult to detect.
These detectors operate locally on the switches,
generating fault reports to the diagnosis manager as needed.
Adding such specialized anomaly detectors
fits within the framework of \name and demonstrates
the extensibility of the approach.
They are part of our implementation (\S\ref{sec:implementation}).
However, we have so far not implemented diagnosis
scripts that use the results of these detectors.

Some faults result in a high rate of route oscillations
or oscillations of the connections between
routing daemons on the switches.
It is relatively simple to modify routing
software to count the number of changes
of the RIB within a given time window.
When the count exceeds a threshold, a potential
fault report is sent to the diagnosis manager.
Excluding periods when significant
network changes are initiated, such reports
can help with diagnosis.
Similarly, it is useful to identify
a high rate of connections and disconnections
of, for example, BGP peers.

The second group of detectors monitor resource usage
on the switches and report resource exhaustion.
We have implemented such detectors of
CPU utilization (load average), memory use,
and use of FIB entries.
 \section{\textit{P\lowercase{odnet}} Details}
\label{appen:podnet}

The Podnet emulation platform is briefly introduced
in \S\ref{sec:emulation}.
This appendix presents additional details regarding
its implementation and fault
injection capabilities.

\begin{figure}[ht]
	\centering
	\includegraphics[width=\columnwidth]{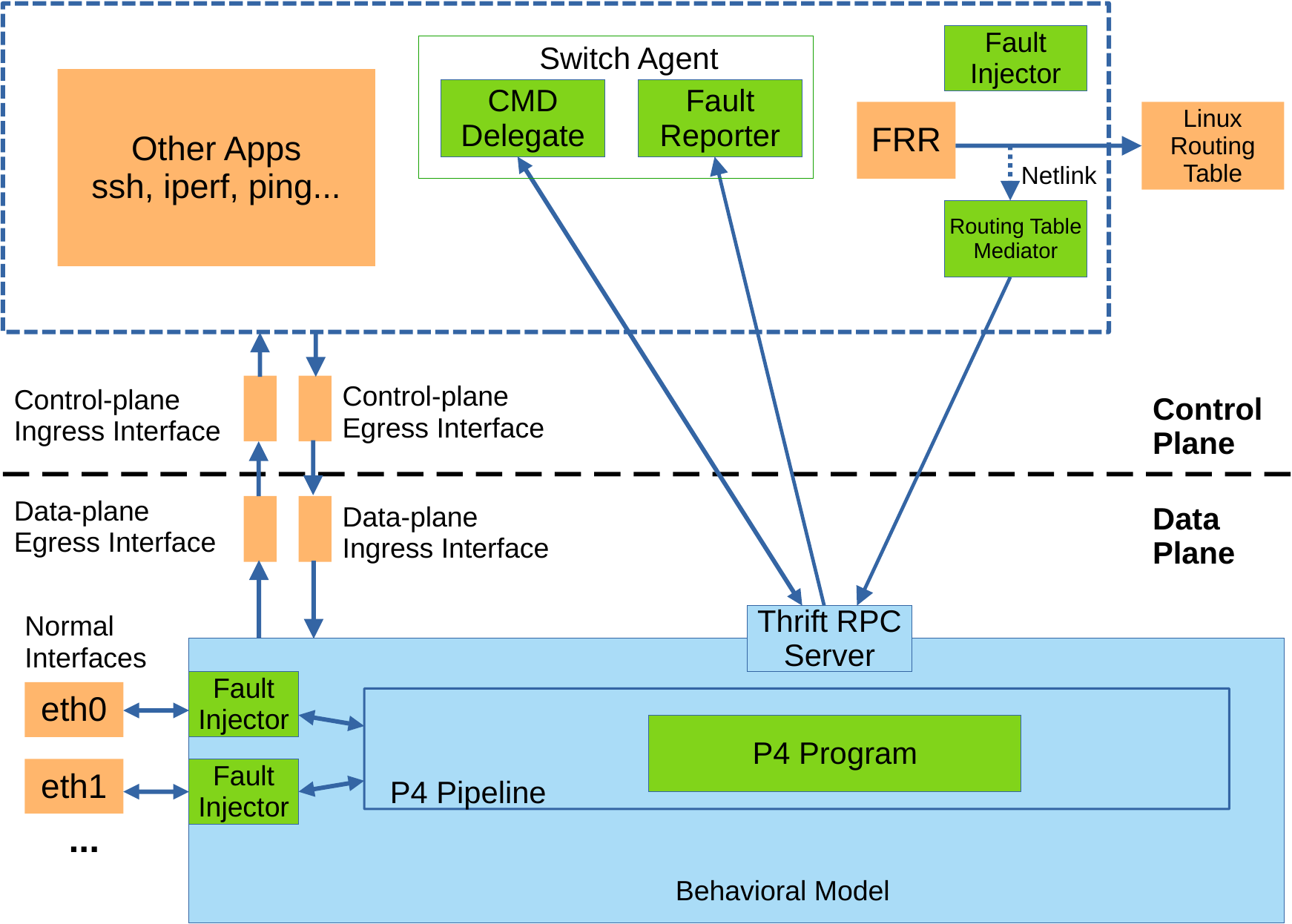}
	\caption{The architecture of an emulated switch.}
	\label{fig:emulated_switch}
\end{figure}

\subsection{Network Emulation}
\label{subsec:emulation_design_impl}

To implement the functionality of a P4 switch,
we use an existing P4 behavioral simulator~\cite{bmv2}.
Each \emuname emulated switch consists of a
P4 simulator instance and an FRR instance.
All the processes that make up an emulated switch
run in a Linux container (managed by Podman~\cite{podman}).
Parts of \emuname are based on
Containernet~\cite{peuster2016containernet},
which, in turn, is based on Mininet~\cite{lantz2010mininet}.
However, neither of these two emulators isolates
each switch in a container.
The architecture of the emulated switch is shown in
Figure~\ref{fig:emulated_switch}.

The implementation of the emulated data plane
requires packets arriving on a virtual network
interface of the container to be directed to
the P4 simulator.
To this end, the P4 simulator captures packets
at the interface using libpcap, while those
packets are blocked
from entering the Linux network stack using iptables.
After packets are processed by the P4 simulator,
they are forwarded to either the local control plane
(data-plane egress interface in Figure~\ref{fig:emulated_switch})
or the interface to the next hop.
Packets from the switch control plane are redirected to
the P4 simulator
(control-plane egress interface in Figure~\ref{fig:emulated_switch})
using iptables.
Thus, these packets are forwarded by
the P4 simulator to the next hop.

The FRR routing software running on the switch container
installs routes into the routing table of the Linux kernel.
FRR does this by sending a Netlink~\cite{Netlink} route change message
to the Linux kernel.
\emuname includes a \textit{routing table mediator}
that monitors route changes in the Linux routing
table via a Netlink socket and installs the
changes to the forwarding table of the P4 simulator
via an RPC interface (Figure~\ref{fig:emulated_switch}). 

\subsection{Fault Injection}
\label{subsec:fault_injection}

To test and evaluate \name, \emuname includes
a fault injector that is able to inject the faults in
our functional fault model.
Below are examples of how specific faults are injected:

\begin{itemize}[nosep,left=0pt]
   \item Silent drop in a switch:
   activate a special P4 stage that drops packets.

   \item Silent drop on a link:
   activate a modification of the packet capturing module in
   the P4 simulator so that it drops packets.

   \item No-forwarding drop caused by inconsistency
   between FIB and RIB:
   activate a special feature of the routing table mediator
   that removes an entry in the FIB.
   
   \item ACL deny drop caused by incorrect
   ACL rule:
   add an ACL rule that denies the particular packets.

   \item BGP neighbor missing caused by a software bug:
   use iptables to block BGP packets.
   
   \looseness=-1
   \item Incorrect BGP advertisement caused
   by a software bug:
   configure iptables to direct packets to an nf-queue monitored
   by a code that uses libnetfilter\_queue.
   This code captures and modifies BGP advertisements that
   contain specified routes.
\end{itemize}
 \begin{table*}[t]
    \centering
    \caption{An instantiated functional model of fault-free and
    faulty switches implementing IPv4 and BGP protocols.}
    \label{tab:instantiated_model}
    {\setlength{\tabcolsep}{3pt}
\setstretch{0.9}
\begin{tabular}{V{2.5} @{ } c @{ } V{2.5}  p{5.8in} V{2.5}}
   \hlineB{2.5}
   \multirow{12}{*}{\rotatebox{90}{Packet Forwarding}} &
   \vspace{-4pt}
   $C_{forward} = \ FIB(Header.IPv4.DstAddr) = Q \ \land \
      Header.IPv4.TTL \neq 0 \ \land $ \\
      & \quad\quad $AccessControl(Header) \neq \text{Deny}\ \land \
        CheckSum(Header.IPv4) = 0$ \\[4pt]
   & $\overline{C_{forward}} = \ FIB(Header.IPv4.DstAddr) = Null \
     \lor \ Header.IPv4.TTL = 0 \ \lor$ \\
     & \quad\quad $AccessControl(Header) = \text{Deny}\ \lor \
     CheckSum(Header.IPv4) \neq 0$ \\
   \clineB{2-2}{1.5}
   & \vspace{-4pt} $EgressPort(C_{forward}, S_{uncongested}) = p,
     p \in Q, Q \subset P$ \\
   & $EgressPort(C_{forward}, S_{congested}) = \text{Null}$ \\
   & $EgressPort(\overline{C_{forward}}, *) = \text{Null}$ \\
   \clineB{2-2}{1.5}
   & \vspace{-4pt} $EgressPort(C_{forward}, S_{uncongested}) = Null$\\
   & $EgressPort(C_{forward}, *) = p, p \notin Q$ \\
   & $EgressPort(\overline{C_{forward}}, *) = p', p' \in P$ \\
\hlineB{2.5}
\multirow{10}{*}{\rotatebox{90}{Packet Transformation}}
& $C = *$ \\
& $Fhdrt(Pkt(C)) = Pkt(C), \text{with} \
     SrcMacAddr \xleftarrow{} EgressPortMac \ \wedge \
     DstMacAddr \xleftarrow{} NextHopMac \ \wedge \ $ \\
& \quad\quad $ TTL \leftarrow TTL_{in} - 1 \ \wedge \ Checksum \leftarrow ChecksumCompute(NewHeader.IPv4)$ \\
& $Fbadt(Pkt(C)) = Pkt(C), \text{with} \ SrcMacAddr \xleftarrow{} Mac' \ \vee DstMacAddr \xleftarrow{} Mac'' \ \vee $ \\
& \quad\quad $ TTL \leftarrow TTL' \ \vee Checksum \leftarrow ChecksumCompute'(NewHeader.IPv4)$ \\
& $\text{where} \ Mac' \neq EgressPortMac, \ Mac'' \neq NextHopMac, \ TTL' \neq TTL_{in} - 1.5, \ ChecksumCompute' \not\equiv ChecksumCompute$\\
\clineB{2-2}{1.5}
& \vspace{-4pt} $NewPkt = Fhdrt(Pkt(C)) \ \land $ \\
& $NewPkt.payload = Pkt(C).payload$ \\
\clineB{2-2}{1.5}
& \vspace{-4pt} $NewPkt = Fbadt(Pkt(C)) \neq Fhdrt(Pkt(C)) \ \lor $ \\
& $NewPkt.payload \neq Pkt(C).payload$ \\
\hlineB{2.5}
\multirow{6.5}{*}{\rotatebox{90}
   {\parbox{\widthof{\fontsize{7}{7}\selectfont Data Plane}}{
      \raggedright\fontsize{7}{7}\selectfont
         Data Plane \\ Table \\ Generation}}}
& \vspace{-4pt} $ControlPlaneTable \leftarrow RIB_{IPv4}\ \&
                    \ ACLConfig$ \\
& $DataPlaneTable \leftarrow FIB_{IPv4} \ \& \ ACL$ \\
\clineB{2-2}{1.5}
& \vspace{-4pt} $\forall Entry \in ControlPlaneTable, \exists Entry' \in DataPlaneTable \ s.t. \ Entry \equiv Entry' \ \land$ \newline
    $\forall Entry \in DataPlaneTable, \exists Entry' \in ControlPlaneTable \ s.t. \ Entry \equiv Entry'$ \\
\clineB{2-2}{1.5}
& \vspace{-4pt} $\exists Entry \in ControlPlaneTable, \forall Entry' \in DataPlaneTable, Entry' \not\equiv Entry \ \lor$ \newline
    $\exists Entry \in DataPlaneTable, \forall Entry' \in ControlPlaneTable, Entry' \not\equiv Entry$ \\
\hlineB{2.5}
\multirow{5.9}{*}{\rotatebox{90}
   {\parbox{\widthof{\fontsize{7}{7}\selectfont Generation}}{
      \raggedright\fontsize{7}{7}\selectfont
         Route \\ Table \\ Generation}}}
& \vspace{-4pt} $RouteTable = RIB_{BGP}$ \\
& $RecvRoutingInformation = RIBin_{BGP}$ \\
& $RouteCompute = BGPRouteCompute$ \\
\clineB{2-2}{1.5}
& \vspace{-4pt} $RouteTable = RouteCompute(Configuration, RecvRoutingInformation)$ \\
\clineB{2-2}{1.5}
& \vspace{-4pt} $RouteTable \neq RouteCompute(Configuration, RecvRoutingInformation)$ \\
\hlineB{2.5}
\multirow{5.5}{*}{\rotatebox{90}
   {\parbox{\widthof{\fontsize{7}{7}\selectfont vertizement}}{
      \raggedright\fontsize{7}{7}\selectfont
         Route Advertisement Reception}}}
& \vspace{-4pt} $RecvRoutingInformation = RIBin_{BGP}$ \\
& $RouteAdvReception = BGPInboundRouteAdvFiltering$ \\
& $InboundRouteAdv = InboundBGPUpdateMsg$ \\
\clineB{2-2}{1.5}
& \vspace{-4pt} $RecvRoutingInformation = RouteAdvReception(Configuration, InboundRouteAdv)$ \\
\clineB{2-2}{1.5}
& \vspace{-4pt} $RecvRoutingInformation \neq RouteAdvReception(Configuration, InboundRouteAdv)$ \\
\hlineB{2.5}
\multirow{5.5}{*}{\rotatebox{90}
   {\parbox{\widthof{\fontsize{7}{7}\selectfont vertizement}}{
      \raggedright\fontsize{7}{7}\selectfont
         Route Advertisement Generation}}}
& \vspace{-4pt} $RouteTable = RIB_{BGP}$ \\
& $RouteAdvGeneration = BGPInboundRouteAdvFiltering$ \\
& $OutboundRouteAdv = OutboundBGPUpdateMsg$ \\
\clineB{2-2}{1.5}
& \vspace{-4pt} $OutboundRouteAdv = RouteAdvGeneration(Configuration, RouteTable)$ \\
\clineB{2-2}{1.5}
& \vspace{-4pt} $OutboundRouteAdv \neq RouteAdvGeneration(Configuration, RouteTable)$ \\
\hlineB{2.5}
\multirow{5.4}{*}{\rotatebox{90}
   {\parbox{\widthof{\fontsize{7}{7}\selectfont with External}}{
      \raggedright\fontsize{7}{7}\selectfont
         Interaction with External Entities}}}
& \vspace{-4pt} $Response \leftarrow TCPResponse/BGPResponse$ \\
& $ProtocolMessage \leftarrow TCPMsg_{SYN-ACK} or BGPOpenMsg/BGPNotificationMsg/BGPRefreshMsg$ \\
\clineB{2-2}{1.5}
& \vspace{-4pt} $Response(InboundMessage, Configuration) = ProtocolMessage$ \\
\clineB{2-2}{1.5}
& \vspace{-4pt} $Response(InboundMessage, Configuration) \neq ProtocolMessage \ \lor$ \newline
    $Response(InboundMessage, Configuration) = Null$\\
\hlineB{2.5}
\end{tabular}
}
 \end{table*}

\section{Model Instantiation for IPv4/BGP}
\label{sec:instantiation}

A high-level functional model of fault-free and faulty switches
is presented in \S\ref{sec:model}.
Table~\ref{tab:instantiated_model} is an instantiation
of this model for IPv4/BGP networks.
As in Table~\ref{tab:formal_model},
the instantiation consists of seven functionality categories.
For each category there are three rows, where the top
row contains the instantiation of the abstract
objects, characteristics, or functions for IPv4/BGP.
The middle row is a definition
of the functionality of a fault-free switch.
The bottom row is a definition of the
functionality of a faulty switch.
The definition of the functionality
of a faulty switch is derived by negating
the functionality of a fault-free switch.
The list of faulty functionalities forms the switch
\textit{fault model}.
  \bibliographystyle{plain}

\end{document}